\documentclass[review,3p,authoryear]{elsarticle}  
 \usepackage{geometry}
\usepackage{framed,multirow}
\usepackage{amssymb}
\usepackage{amsmath}
\usepackage{latexsym}

\usepackage{url}
\usepackage{xcolor}
\usepackage{hyperref}
\usepackage{caption}
\usepackage{subcaption}
\usepackage{graphicx}
\usepackage[figuresright]{rotating}
\usepackage{tabularray}
\usepackage{pdflscape}

\usepackage{tikz}
\newcommand*{\tikzbullet}[2]{%
  \setbox0=\hbox{\strut}%
  \begin{tikzpicture}
    \useasboundingbox (-.35em,0) rectangle (.35em,\ht0);
    \filldraw[draw=#1,fill=#2] (0,0.5\ht0) circle[radius=.35em];
  \end{tikzpicture}%
}

\definecolor{amaranth}{rgb}{0.9, 0.17, 0.31}

\definecolor{atomictangerine}{rgb}{1.0, 0.6, 0.4}

\definecolor{Darkcyan}{rgb}{0,0.573,0.573}

\definecolor{byzantine}{rgb}{0.74, 0.2, 0.64}

\definecolor{lightpink}{rgb}{1,0.427,0.714}

\definecolor{aurometalsaurus}{rgb}{0.43, 0.5, 0.5}

\definecolor{palepink}{rgb}{1,0.714,0.859}

\definecolor{lightviolet}{rgb}{0.714,0.427,1}

\definecolor{pureblue}{rgb}{0,0.427,0.859}

\definecolor{limegreen}{rgb}{0.141,1,0.141}

\definecolor{lightyellow}{rgb}{1,1,0.427}

\definecolor{paleblue}{rgb}{0.714,0.859,1}

\definecolor{aquamarine}{rgb}{0.5, 1.0, 0.83}

\definecolor{pureorange}{rgb}{0.859,0.427,0}

\definecolor{lightblue}{rgb}{0.427,0.714,1}

\definecolor{newcolor}{rgb}{.8,.349,.1}

\begin{document}


\begin{frontmatter}

\title{The Brain Tumor Sequence Registration (BraTS-Reg) Challenge: Establishing Correspondence Between Pre-Operative and Follow-up MRI Scans of Diffuse Glioma Patients}




 

\author[iu,first]{Bhakti Baheti\textsuperscript{*,\#,}}   
\author[washuese,gehc,first]{Satrajit Chakrabarty\textsuperscript{*,}} 
\author[santaclara]{Hamed Akbari\textsuperscript{*,}} 
\author[upennrad]{Michel Bilello\textsuperscript{*,}}  
\author[tumrad]{Benedikt Wiestler\textsuperscript{*,}} 
\author[tumrad]{Julian Schwarting\textsuperscript{*,}}  
\author[duke]{Evan Calabrese\textsuperscript{*,}} 
\author[ucsd]{Jeffrey Rudie\textsuperscript{*,}}   
\author[washurad]{Syed Abidi\textsuperscript{*,}}  
\author[washurad]{Mina Mousa\textsuperscript{*,}}  
\author[ucsf]{Javier Villanueva-Meyer\textsuperscript{*,}} 
\author[ucsf]{Brandon K.K. Fields\textsuperscript{*,}}
\author[helm,tuminfo,tumrad] {Florian Kofler\textsuperscript{*,}} 
\author[cbica,taki] {Russell Takeshi Shinohara\textsuperscript{*,}}

\author[mgh,vxm-1,UC-london]{Juan Eugenio Iglesias\textsuperscript{*,}}

\author[cwmok]{Tony C. W. Mok}  
\author[cwmok]{Albert C. S. Chung} 

\author[AGH-1,AGH-2]{Marek Wodzinski}
\author[AGH-1,AGH-2]{Artur Jurgas}
\author[AGH-1,AGH-3]{Niccol\`{o} Marini}
\author[AGH-1,AGH-4]{Manfredo Atzori}
\author[AGH-1,AGH-5]{Henning M\"{u}ller}

\author[UZL]{Christoph Großbr\"{o}hmer}
\author[UZL]{Hanna Siebert}
\author[UZL]{Lasse Hansen}
\author[UZL]{Mattias P. Heinrich}

\author[MeVIS-1,MeVIS-2]{Luca Canalini}
\author[MeVIS-1]{Jan Klein}
\author[MeVIS-1]{Annika Gerken}
\author[MeVIS-3]{Stefan Heldmann}
\author[MeVIS-3,MeVIS-4]{Alessa Hering}
\author[MeVIS-1,MeVIS-2]{Horst K. Hahn}

\author[BDAV-USYD-1]{Mingyuan Meng}
\author[BDAV-USYD-1]{Lei Bi}
\author[BDAV-USYD-1,BDAV-USYD-2]{Dagan Feng}
\author[BDAV-USYD-1]{Jinman Kim}

\author[camed-1,camed-2,camed-3]{Ramy A. Zeineldin}
\author[camed-2]{Mohamed E. Karar}
\author[camed-3]{Franziska Mathis-Ullrich}
\author[camed-1]{Oliver Burgert}

\author[Kurtlab]{Javid Abderezaei}
\author[Kurtlab]{Aymeric Pionteck}
\author[Kurtlab]{Agamdeep Chopra}
\author[Kurtlab]{Mehmet Kurt}

\author[YKW]{Kewei Yan}
\author[YKW]{Yonghong Yan}

\author[SuperX]{Zhe Tang}
\author[SuperX]{Jianqiang Ma}

\author[medal-1]{Sahar Almahfouz Nasser}
\author[medal-1]{Nikhil Cherian Kurian}
\author[medal-1]{Mohit Meena}
\author[medal-2]{Saqib Shamsi}
\author[medal-1]{Amit Sethi}

\author[ANTs-1]{Nicholas J. Tustison}
\author[ANTs-1]{Brian B. Avants}
\author[ANTs-2]{Philip Cook}
\author[ANTs-2]{James C. Gee}

\author[gradicon-1]{Lin Tian}
\author[gradicon-1]{Hastings Greer}
\author[gradicon-1]{Marc Niethammer}

\author[vxm-1,vxm-2,vxm-3]{Andrew Hoopes}
\author[vxm-2,vxm-3,vxm-4,vxm-5]{Malte Hoffmann}
\author[vxm-1,vxm-2,vxm-3,vxm-4]{Adrian V.\ Dalca}

\author[mics-1]{Stergios Christodoulidis}
\author[mics-1]{Th\'eo Estiene}
\author[mics-1]{Maria Vakalopoulou} 
\author[mics-2]{Nikos Paragios}

\author[washurad]{Daniel S. Marcus\textsuperscript{*,}} 
\author[cbica,upennrad]{Christos Davatzikos\textsuperscript{*,}} 
\author[washurad,washui2,senior]{Aristeidis Sotiras\textsuperscript{*,\#,}} 
\author[uzh,tuminfo,senior]{Bjoern Menze\textsuperscript{*,\#,}}  
\author[iu,iurad,iunsg,iucs,senior]{Spyridon Bakas\textsuperscript{*,\#,}}  
\author[uzh,tuminfo,regnuk,regdiz,senior]{Diana Waldmannstetter\textsuperscript{*,\#,}}   

\vspace{2mm}

\address[iu]{Division of Computational Pathology, Department of Pathology and Laboratory Medicine, Indiana University School of Medicine, Indianapolis, IN, USA}

\address[washuese]{Department of Electrical and Systems Engineering, Washington University in St.
Louis, St. Louis, MO, USA}
\address[gehc]{
GE HealthCare, San Ramon, CA, USA
}
\address[cbica]{Center for Artificial Intelligence and Data Science for Integrated Diagnostics (AI2D) and Center for Biomedical Image Computing and Analytics (CBICA), University of Pennsylvania, Philadelphia, PA, USA}
\address[upennrad]{Department of Radiology, Perelman School of Medicine, University of Pennsylvania, Philadelphia, PA, USA}

\address[washurad]{Mallinckrodt Institute of Radiology, Washington University School of Medicine, St.
Louis, MO, USA}
\address[washui2]{Institute for Informatics, Data Science \& Biostatistics, Washington University School of Medicine, St. Louis, MO, USA}

\address[santaclara]{Department of Bioengineering, Santa Clara University, Santa Clara, CA, USA} 

\address[uzh]{Department of Quantitative Biomedicine, University of Zurich, Zurich, Switzerland}
\address[tuminfo]{Department of Computer Science, TUM School of Computation, Information and Technology, Technical University of Munich, Munich, Germany}
\address[tumrad]{Department of Diagnostic and Interventional Neuroradiology, School of Medicine, Klinikum rechts der Isar, Technical University of Munich, Munich, Germany}
\address[regnuk]{Department of Nuclear Medicine, University Hospital Regensburg, Regensburg, Germany}
\address[regdiz]{Medical Data Integration Center (MEDIZUKR), University Hospital Regensburg, Regensburg, Germany}

\address[ucsf]{Center  for Intelligent  Imaging (ci2), Department of Radiology \& Biomedical Imaging, University of California, San Francisco, CA, USA}

\address[duke]{Department of Radiology, Duke University Medical Center, Durham, NC, USA} 

\address[ucsd]{Department of Radiology, University of California, San Diego, CA, USA} 

\address[helm]{Helmholtz AI, Helmholtz Munich, Neuherberg, Germany}

\address[taki]{Penn Statistics in Imaging and Visualization Center, Department of Biostatistics, Epidemiology, and Informatics, University of Pennsylvania, Philadelphia, USA}

\address[mgh]{Massachusetts General Hospital, Harvard Medical School, USA}
\address[UC-london]{Centre for Medical Imaging Computing, University College London, United Kingdom}
\address[cwmok]{The Hong Kong University of Science and Technology, Hong Kong, China}

\address[AGH-1]{University of Applied Sciences Western Switzerland and Information Systems Institute, Sierre, Switzerland}
\address[AGH-2]{AGH University of Krakow, Department of Measurement and Electronics, Krakow, Poland} 
\address[AGH-3]{Department of Computer Science, University of Geneva, Geneva, Switzerland}
\address[AGH-4]{Department of Neuroscience, University of Padova, Padova, Italy}
\address[AGH-5]{Medical Faculty, University of Geneva, Geneva, Switzerland}

\address[UZL]{Institute of Medical Informatics, Universität zu Lübeck, Lübeck, Germany}

\address[MeVIS-1]{Fraunhofer MEVIS, Institute for Digital Medicine, Bremen, Germany}
\address[MeVIS-2]{University of Bremen, Bremen, Germany}
\address[MeVIS-3]{Fraunhofer MEVIS, Institute for Digital Medicine, Lübeck, Germany}
\address[MeVIS-4]{Diagnostic Image Analysis Group, Radboud University Medical Center, Nijmegen, The Netherlands}

\address[BDAV-USYD-1]{School of Computer Science, The University of Sydney, Sydney, Australia}
\address[BDAV-USYD-2]{Med-X Research Institute, Shanghai Jiao Tong University, Shanghai, China}

\address[YKW]{University of North Carolina at Charlotte, Charlotte NC 28223, USA}

\address[SuperX]{Keya Medical Technology Co. Ltd, Beijing, China}

\address[medal-1]{Indian Institute of Technology, Bombay}
\address[medal-2]{Whirlpool, Pune, India}

\address[camed-1]{Research Group Computer Assisted Medicine (CaMed), Reutlingen University, Germany}
\address[camed-2]{Faculty of Electronic Engineering (FEE), Menoufia University, Egypt}
\address[camed-3] {Department Artificial Intelligence in Biomedical Engineering, Friedrich-Alexander-University Erlangen-Nürnberg, Germany} 

\address[Kurtlab]{Department of Mechanical Engineering, University of Washington, Seattle, USA}

\address[ANTs-1]{Department of Radiology and Medical Imaging, University of Virginia}
\address[ANTs-2]{Penn Image Computing and Science Lab, Department of Radiology, Perelman School of Medicine at the University of Pennsylvania, Philadelphia, PA, USA}

\address[vxm-1]{Computer Science and Artificial Intelligence Laboratory, Massachusetts Institute of Technology, 32 Vassar St, Cambridge, MA 02139, USA}
\address[vxm-2]{Athinoula A.\ Martinos Center for Biomedical Imaging, 149 13\textsuperscript{th} St, Charlestown, MA 02129, USA}
\address[vxm-3]{Department of Radiology, Massachusetts General Hospital, 55 Fruit St, Boston, MA 02114, USA}
\address[vxm-4]{Department of Radiology, Harvard Medical School, 25 Shattuck St, Boston, MA 02115, USA}
\address[vxm-5]{Harvard-MIT Health Sciences and Technology, Massachusetts Institute of Technology, 77 Massachusetts Ave, Cambridge, MA 02139, USA}

\address[gradicon-1]{University of North Carolina at Chapel Hill}

\address[mics-1]{CentraleSup\'elec, 3 Rue Joliot Curie, 91190 Gif-sur-Yvette, France}
\address[mics-2]{Therapanacea, 7 bis Bd Bourdon, 75004 Paris, France}

\address[iurad]{Department of Radiology and Imaging Sciences, Indiana University School of Medicine, Indianapolis, IN, USA}
\address[iunsg]{Department of Neurological Surgery, Indiana University School of Medicine, Indianapolis, IN, USA}
\address[iucs]{Department of Computer Science, Luddy School of Informatics Computing and Engineering, Indiana University, Indianapolis, IN, USA}

\address[first]{Equally contributing first authors}
\address[senior]{Equally contributing senior authors}
\address{* People involved in organization of the BraTS-Reg challenge}
\address{\# Corresponding authors: bvbaheti@iu.edu, aristeidis.sotiras@wustl.edu, bjoern.menze@uzh.ch, spbakas@iu.edu, diana.waldmannstetter@tum.de}

\begin{abstract}
Registration of longitudinal brain Magnetic Resonance Imaging (MRI) scans containing pathologies is challenging due to dramatic changes in tissue appearance. Although there has been considerable progress in developing general-purpose medical image registration techniques, they have not yet attained the requisite precision and reliability for this task, highlighting its inherent complexity. Here we describe the Brain Tumor Sequence Registration (BraTS-Reg) challenge, as the first public benchmark environment for deformable registration algorithms focusing on estimating correspondences between pre-operative and follow-up scans of the same patient diagnosed with a diffuse brain glioma. The challenge was conducted in conjunction with both the IEEE International Symposium on Biomedical Imaging (ISBI) 2022 and the International Conference on Medical Image Computing and Computer-Assisted Intervention (MICCAI) 2022. The BraTS-Reg data comprise de-identified multi-institutional multi-parametric MRI (mpMRI) scans, curated for size and resolution according to a canonical anatomical template, and divided into training, validation, and testing sets. Clinical experts annotated ground truth (GT) landmark points of anatomical locations distinct across the temporal domain. The training data with their GT annotations, were publicly released to enable the development of registration algorithms. The validation data, without their GT annotations, were also released to allow for algorithmic evaluation prior to the testing phase, which only allowed submission of containerized algorithms for evaluation on hidden hold-out testing data. Quantitative evaluation and ranking was based on the Median Euclidean Error (MEE), Robustness, and the determinant of the Jacobian of the displacement field. The top-ranked methodologies yielded similar performance across all evaluation metrics and shared several methodological commonalities, including pre-alignment, deep neural networks, inverse consistency analysis, and test-time instance optimization per-case basis as a post-processing step. The top-ranked method attained the MEE at or below that of the inter-rater variability for approximately 60\% of the evaluated landmarks, underscoring the scope for further accuracy and robustness improvements, especially relative to human experts. The aim of BraTS-Reg is to continue to serve as an active resource for research, with the data and online evaluation tools accessible at \url{https://bratsreg.github.io/}. 
\end{abstract}

\begin{keyword}
BraTS-Reg \sep Registration \sep Glioma \sep MRI \sep Longitudinal \sep Diffuse \sep glioma \sep Glioblastoma
\end{keyword}

\end{frontmatter}


\section{Introduction}
Registration is a fundamental problem in medical image analysis \citep{sotiras2013deformable, ou2014comparative} that aims to find spatial correspondences between two images and align them for various downstream applications. 
Accurate longitudinal image registration between pre-operative and follow-up scans is particularly crucial for patients with brain tumors. 
Such registration can aid in analyzing the characteristics of healthy tissue, potentially identifying tumor recurrence \citep{han2020deep}. 
Furthermore, it can enhance our understanding of underlying pathophysiological processes and improve treatment response assessment.
Diffuse glioma, and specifically Isocitrate Dehydrogenase (IDH)-wildtype glioblastoma as per the World Health Organization (WHO) classification of tumors of the central nervous system (CNS WHO grade 4)  \citep{who2021classification}, is the most common and aggressive malignant adult brain tumor that heavily and heterogeneously infiltrates and deforms its surrounding brain tissue. 
Finding imaging signatures in the pre-operative setting that can predict tumor infiltration and subsequent tumor recurrence is very important in the treatment and management of brain diffuse glioma patients \citep{akbari2016imaging}, as it could influence treatment decisions even at baseline patient evaluations \citep{kwon2015estimating}.

\begin{figure}[!t]
\centering
    {\includegraphics[width=12cm]{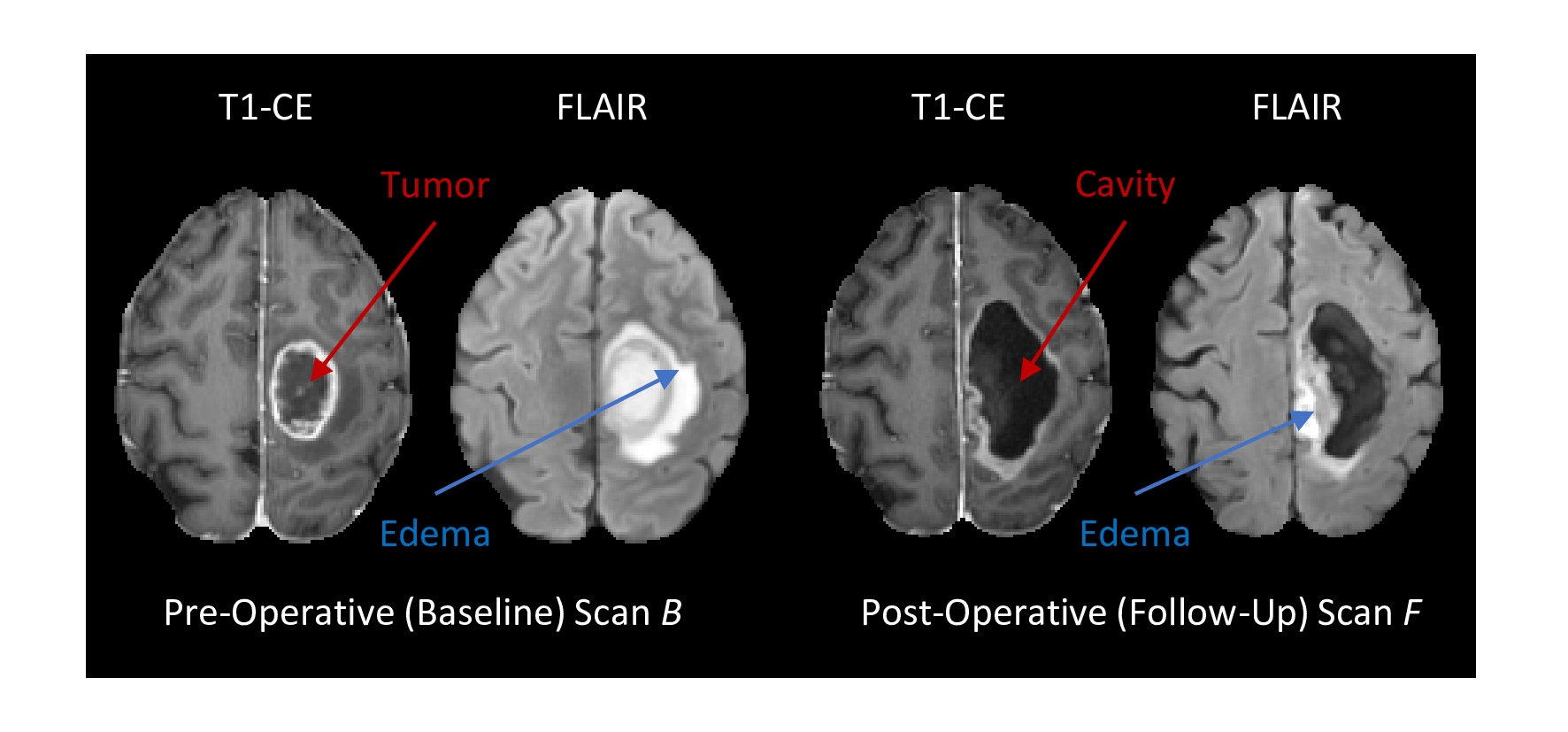}}
    \caption {Example of a pre-operative baseline and its corresponding follow-up MRI scan. The contrast-enhanced T1-weighted (T1-CE), and the T2 Fluid Attenuated Inversion Recovery (FLAIR) baseline scan clearly show the tumor and the edema, respectively. Similarly, the T1-CE follow-up scan shows the resection cavity, whereas the edema is visible in the FLAIR scan.}
    \label{Baseline_followup_example}
\end{figure}

The registration between pre-operative and follow-up multi-parametric Magnetic Resonance Imaging (mpMRI) scans of patients with diffuse glioma is important yet challenging due to i) large deformations in the brain tissue, caused by the tumor’s mass effect, ii) missing correspondences between the apparent tumor in the pre-operative baseline scan and the resection cavity in the follow-up scans, and iii) inconsistent intensity profiles between the acquired scans, as MRI acquisition does not depend on standardized units unlike, for example, Computed Tomography (CT) that depends on Hounsfield units. 
Fig.~\ref{Baseline_followup_example} shows a baseline pre-operative and follow-up scan of the same glioma patient, with the resection cavity, tumor, and peritumoral edematous regions marked. 
The peritumoral tissue labeled as `edema' in the pre-operative scan, is known to also include infiltrated tumor cells that may lead to recurring tumors apparent in follow-up scans. 
Thus, corresponding brain tissue regions can have highly heterogeneous intensity profiles across the same brain. 
This raises the need for the development of accurate deformable registration algorithms to establish spatial correspondences between the pre-operative and follow-up mpMRI brain scans. 
This would facilitate mapping follow-up information onto baseline scans to elucidate imaging signatures that can be used for detection of recurrence in future unseen cases \citep{akbari2020histopathology}. 
Such algorithms must be able to account for the large deformations in the pre-operative scan due to the tumor’s mass effect, as well as their relaxation in follow-up scans after tumor resection. 
Additional complexities arise from alterations introduced by tumor resection, follow-up radiation treatment, and, in many cases, local tumor progression. 
Addressing these challenges remains open despite several decades of research in image registration, even specific to this problem \citep{kwon2013portr,kwon2015estimating,han2018patient,ou2014comparative}.
Prior work has typically explored modifying standard registration algorithms by including cost-function masking (i.e., masking out the tumor or cavity areas from the alignment process) and interpolating displacements of the underlying tissue from nearby healthy areas. 
However, no comprehensive analysis of these approaches is currently available. 
Importantly, there is a lack of efforts to measure local alignment errors comprehensively. 

This paper describes the design and outcomes of the Brain Tumor Sequence Registration (BraTS-Reg) challenge, which provided a substantial collection of retrospective multi-institutional mpMRI data, representing patient populations with distinct demographics. 
The primary aim of BraTS-Reg was to establish a public benchmark environment for deformable registration algorithms. 
The objectives of the challenge were to: 
i) identify the most effective registration algorithm for the specific task, and 
ii) establishing a quantitative baseline by assessing the accuracy of current state-of-the-art algorithms.
The rest of the manuscript provides an overview of prior work in this field, followed by a complete summary of the BraTS-Reg challenge conducted in conjunction with both the IEEE International Symposium on Biomedical Imaging (ISBI) 2022 and the International Conference on Medical Image Computing and Computer-Assisted Intervention (MICCAI) 2022. 
It describes all the details including, but not limited to, the data description, the evaluation approach, the summary of participating methods, and the meta-analysis of the obtained results. 

\section{Prior Work}
\subsection{Approaches for Baseline-to-follow-up Registration} 
Few approaches have been proposed to tackle the task of image registration between pre-operative and follow-up mpMRI scans.
In 2013, Kwon et al. \citep{kwon2013portr} developed \emph{PORTR}, a registration approach combining registration and segmentation through an expectation-maximization optimization approach coupled with a biophysical tumor growth modeling framework. 
They continued their research using patient-specific templates to estimate brain anatomy prior to tumor evolution \citep{kwon2015estimating}. \citep{chen2015deformable} coupled image registration and automatic detection of regions with missing correspondences and \citep{feng2015region} modeled region correspondences between images using graph matching. 
An approach for combining image registration with reconstruction was presented by \citep{han2018patient} and a method for finite element biomechanical modeling for registration using image-guided neurosurgery was developed in \citep{drakopoulos2016accurate}. 
Biomechanical models can benefit brain deformation analysis as well as tumor progression evaluation, as Lipkova et al. \citep{lipkova2022modelling} and Ezhov et al. \citep{ezhov2023learn} demonstrated. 
\citep{waldmannstetter2020reinforced} utilized deep reinforcement learning for re-detecting landmarks in pre- and post-operative brain scans. 
There has also been some development in the field of deformable registration of images with pathologies in general, including the simulation of tumor mass effect like in \citep{zacharaki2008orbit}, where patient-specific brain anatomy is imitated and in \citep{han2020deep}, where a deep network is utilized to reconstruct images with pathologies by disentangling the tumor mass effect from the reconstruction of quasi-normal images. 
Integrating image registration with biophysical modeling is a field with numerous advances \citep{kwon2013portr,kwon2015estimating,scheufele2019coupling}. 
Nevertheless, there is still research required in this field, especially when it comes to fully automatic registration algorithms. 
Moreover, most of the referenced methods have been tested on limited amounts of local data and have not been benchmarked to demonstrate stable performance across common multi-institutional datasets.

\subsection{Related Challenges and Benchmarks}
In the last few years, multiple challenges and benchmarks related to image registration, or brain tumors, have taken place. Starting in 2012 and ongoing to this date, the \emph{Brain Tumor Segmentation (BraTS) Benchmark} \citep{menze2014multimodal,bakas2017advancing,bakas2018identifying,bakas2017segmentationGBM,bakas2017segmentationLGG,baid2021rsna} has shown a remarkable impact on the development of algorithms for the segmentation of brain tumors. 
For a subset of the training and testing data of BraTS, several scans are available acquired at pre- and post-operative timepoints. In the field of image registration, multiple challenges and benchmarks have been designed. 
\emph{Learn2Reg} \citep{hering2022learn2reg} yearly organizes data for several registration tasks. \emph{The Continuous Registration Challenge} \citep{marstal2019continuous} provides a platform for benchmarking registration methods for the registration of lung CT and brain MRI scans in a fully automated way. 
Various other challenges exist, addressing the evaluation of MRI-to-ultrasound registration methods for brain shift correction (\emph{CuRIOUS}) \citep{xiao2019evaluation}, the automated registration of histological images (\emph{ANHIR}) \citep{borovec2020anhir}, the automated registration of breast cancer tissue (\emph{ACROBAT}) \citep{weitzacrobat}, and the evaluation of registration methods on thoracic CT (\emph{EMPIRE10}) \citep{murphy2011evaluation}. 

\subsection{Existing Datasets for Deformable Registration and Baseline-to-follow-up Registration}
Several datasets suitable for the task of deformable, and baseline-to-follow-up registration, are publicly available.

\emph{QIN GBM Treatment Response}: MRI data from 54 newly diagnosed glioblastoma patients, including pre- and post-operative images. Post-operative scans were acquired after surgery, but prior to therapy start \citep{prah2015repeatability, mamonov2016data,Mamonov2016-rd}.

\emph{IvyGap}: 390 studies for 39 patients including pre-operative, post-operative and follow-up mpMRI scans. Genomic information and digitized histopathology tissue sections are also provided \citep{puchalski2018anatomic,Shah2016-ii}.

\emph{UPenn-GBM}: MRI data from 60 patients with pre-operative and follow-up scans, which were acquired after a second resection. 
The dataset includes information on genomics, radiomics, and digitized tissue sections for a comprehensive analysis \citep{bakas2022university,Bakas2021-ez}.

\emph{LUMIERE}: Longitudinal MRI data of 91 glioblastoma patients with pre-operative and follow-up scans. Additional provided information includes expert ratings according to the RANO guidelines, patient age at the time of diagnosis, sex, overall survival time, tumor segmentations, radiomics, and pathology information \citep{suter2022lumiere}.

Nevertheless, the amount of patient data including pre- and post-operative scans offered by each of these datasets is limited (\textless 100 patients). 
This is particularly limiting for deep learning algorithms, which typically require a larger dataset for effective training and evaluations. 
Importantly, none of the listed datasets include expert landmark annotations. 

\section{Challenge Description}
\subsection{Overview}
We organized the first BraTS-Reg challenge focusing on estimating correspondences between baseline pre-operative and follow-up scans of patients with brain gliomas and intended to establish a benchmark environment for deformable registration algorithms. 
The challenge was first organized in March 2022, and conducted as a virtual event in conjunction with the IEEE ISBI. 
It was later held as an in-person event in September 2022, in conjunction with the MICCAI conference in Singapore. 
Interested participants were required to register on the online evaluation portal, which was also previously utilized as the evaluation platform for the BraTS challenge from 2017 to 2021 \citep{menze2014multimodal, bakas2017advancing, bakas2018identifying, baid2021rsna, bakas2017segmentationGBM, bakas2017segmentationLGG}. 
Upon registration, participants were granted access to i) training data including GT annotations, and ii) validation data without GT annotations. 
They were also provided with the necessary resources for the quantitative evaluation of their containerized algorithms. 
The portal automatically computed and returned detailed case-wise performance scores to participants for their submissions. 
We further maintained an unranked live leaderboard \footnote{\url{https://www.cbica.upenn.edu/BraTSReg22/}} for the training and validation phases. 
Participants were given 3.5 months from the challenge start date to submit the containerized algorithm and short paper describing their method and results. 
The organizers evaluated the containers on testing data and reviewed the paper to ensure it contained sufficient details required to understand and reproduce the algorithm within 30 days.  



The participants who secured top rankings in the testing phase of both the ISBI and MICCAI challenges were invited to orally present their methods. 
The final challenge rankings were officially announced during the respective conferences. 
Participants from both the ISBI and MICCAI challenges who submitted functioning containerized algorithms were given the opportunity to extend their papers. 
Accepted papers, after a double-blind review process, were published in Springer Lecture Notes in Computer Science (LNCS) \citep{mok2022robust,wodzinski2022unsupervised,grossbrohmer2022employing,canalini2022iterative,meng2022brain,almahfouz2022wssamnet,abderezaei20223d,zeineldin2022self, yan2022applying}.

\subsection{Multi-institutional Data Sources}
The mpMRI data utilized in the BraTS-Reg challenge consisted of curated and pre-processed retrospective multi-institutional data obtained from routine clinical practice. Specifically, the BraTS-Reg data comprise 259 diffuse glioma patients from our affiliated institutions and from the publicly available The Cancer Imaging Archive (TCIA) \citep{TCIA} collections of: 
i) TCGA-GBM \citep{scarpace2016radiology}, ii) TCGA-LGG \citep{pedano2016radiology}, iii) IvyGAP \citep{ivygap1_puchalski2018anatomic, ivygap2_shah2016data}, and iv) CPTAC-GBM \citep{CPTAC_GBM, wang2021proteogenomic}. 
For the private institutional datasets, the protocol for releasing the data was approved by the institutional review board of the contributing institutions. 
The complete dataset consisted of 259 pairs of pre-operative baseline and follow-up brain mpMRI scans, with each pair corresponding to the same patient diagnosed and treated for an adult diffuse glioma (WHO CNS grades 3-4). 
The specific mpMRI sequences at each time-point were i) native T1-weighted (T1), ii) contrast-enhanced T1 (T1-CE), iii) T2-weighted (T2), and iv) T2 Fluid Attenuated Inversion Recovery (FLAIR).
\begin{figure}[!t]
    \centering
        {\includegraphics[width=10cm]{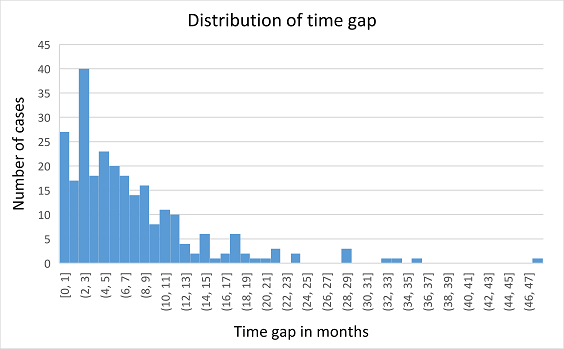}}
        \caption {Distribution of time gap between baseline and followup scan over the complete data used in BraTS-Reg challenge. Each bin of the histogram indicates one month starting from 0 to 48.}
        \label{BraTSReg timegap}
    \end{figure}
    
Following close coordination with the clinical experts of the organizing committee (H.A., M.B., B.W., J.S., E.C., J.R., S.A., M.M.), the time-window between the two paired scans of each patient was decided to be selected such that i) the scans of the two time-points had sufficient apparent tissue deformations, and ii) confounding effects of surgically induced contrast enhancement \citep{albert1994early, wen2010updated} were avoided. 
Therefore, after a thorough visual assessment, the identified follow-up data had to be at least 27 days after the initial surgery. 
Thus, the time-window between all pairs of baseline and follow-up mpMRI scans was in the range of 27 days -- 48 months (Fig. \ref{BraTSReg timegap}).

\subsection{Data Preparation}
The complete multi-institutional dataset consisted of scans acquired under routine clinical conditions, and hence reflected very heterogeneous acquisition equipment and protocols, resulting in differences in the image properties. To ensure consistency and keep the challenge focused on the registration problem, all included scans were pre-processed and provided to the participants in the Neuroimaging Informatics Technology Initiative (NIfTI) file format \citep{nifti}. 
Specifically, three steps of pre-processing were performed for the complete challenge dataset, in line with the BraTS pre-processing protocol \citep{menze2014multimodal,bakas2017advancing,bakas2018identifying,baid2021rsna}: 
i) all scans were first re-oriented into the Left-Post-Superior (LPS) coordinate system and rigidly co-registered to the same canonical anatomical space (i.e., the SRI24 atlas \citep{SRI_rohlfing2010sri24}). 
ii) all scans were then resampled to the same isotropic resolution (i.e., $1 mm^{3}$). 
iii) subsequently, brain extraction was performed to remove non-cerebral tissues like the skull, scalp, and dura from scans \citep{thakur2020brain}. 
Depending on the data source, these steps have been performed using different tools\citep{thakur2020brain,captk,kofler2020brats,chakrabarty2023integrative}, all resulting in the same image format. 
Rigid registration was performed using either ``Greedy''\footnote{\url{github.com/pyushkevich/greedy}} \citep{yushkevich2016fast}, a CPU-based C++ implementation of the greedy diffeomorphic registration algorithm \citep{joshi2004unbiased}, or ANTs \citep{avants2011reproducible}, depending on the institutional source. 
Greedy shares multiple concepts and implementation strategies within the SyN tool in the ANTs package but focuses on computational efficiency. 
Brain extraction was performed using Brain Mask Generator (BrainMaGe\footnote{\url{github.com/CBICA/BrainMaGe}}) \citep{thakur2019skull, thakur2020brain} or HD-BET \citep{isensee2019automated}. 
BrainMaGe is based on a deep learning segmentation architecture (namely U-Net \citep{ronneberger2015u}) and uses a novel training strategy introducing the brain shape as a prior and hence allowing it to be agnostic to the input MRI sequence. 
HD-BET is another deep learning-based brain extraction method trained on glioma patients. 
Notably, the complete pre-processing pipeline has been incorporated in the BraTS Toolkit \citep{kofler2020brats}, the Cancer Imaging Phenomics Toolkit (CaPTk\footnote{\url{www.cbica.upenn.edu/captk}} \citep{captk, fathi2020cancer, captk_2, captk_3}), and the Integrative Imaging Informatics for Cancer Research: Workflow Automation for Neuro-Oncology (I3CR-WANO\footnote{\url{https://github.com/satrajitgithub/NRG_AI_NeuroOnco_preproc}}) framework \citep{chakrabarty2023integrative}, which were actually used for the pre-processing of the provided sequences.

\subsection{Landmark Annotation Protocol}

\begin{figure}
\centering
    \includegraphics[width=10cm]{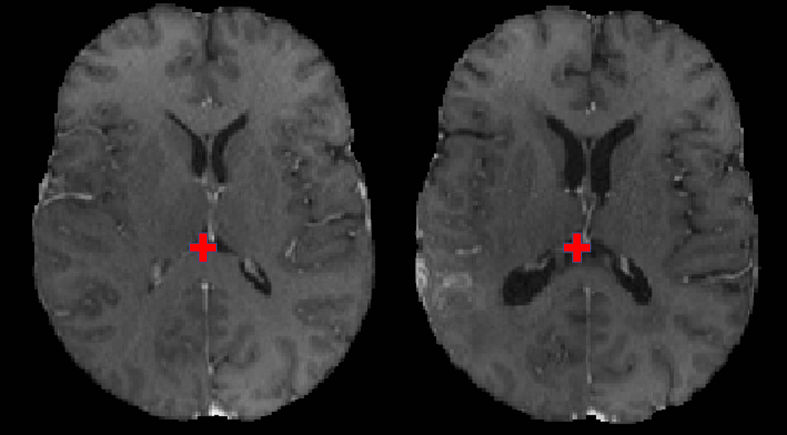}
    \caption{An example of corresponding pair of landmark point in Baseline (left) and Follow-up (Right) superimposed on t1ce scan for visualization}
    \label{Annotation_example}
\end{figure}

The clinical experts of our team (H.A., M.B., B.W., J.S., E.C., J.R., S.A., M.M.) were provided with the segmentation of the tumor core (i.e., the potentially resectable region) and with specific annotation instructions. 
Specifically, for each pre-operative scan, an expert placed $\chi$ number of landmarks near the tumor (within $30 mm$) and $\psi$ number of landmarks far from the tumor (beyond $30 mm$). 
Subsequently, the expert identified the corresponding landmarks in the post-operative follow-up scan. For the data used in the longitudinal analyses, the corresponding landmarks were identified by the experts in the second follow-up scans as well.
The landmarks were defined on anatomically distinct locations, such as blood vessel bifurcations and anatomical landmarks of the midline of the brain.
Fig. \ref{Annotation_example} shows a sample landmark point marked in the baseline scan along with the location of its corresponding landmark in the follow-up scan. 
The total number of landmarks ($\chi$+$\psi$) varied for each case between $6$ and $50$. 
The annotators were given the flexibility to use their preferred tool (including MIPAV \citep{mcauliffe2001medical}, CaPTk \citep{captk}, and ITK-SNAP \citep{yushkevich2016itk}) for making the annotations. 
Finally, the coordinates of the landmark locations were saved into a common comma-separated value (.csv) format.

\subsection{Inter-Rater Annotation Variability}\label{IR_sec}
\begin{figure}[!htbp]
         \centering
         \begin{subfigure}[b]{0.49\textwidth}
             {\includegraphics[width=\textwidth]{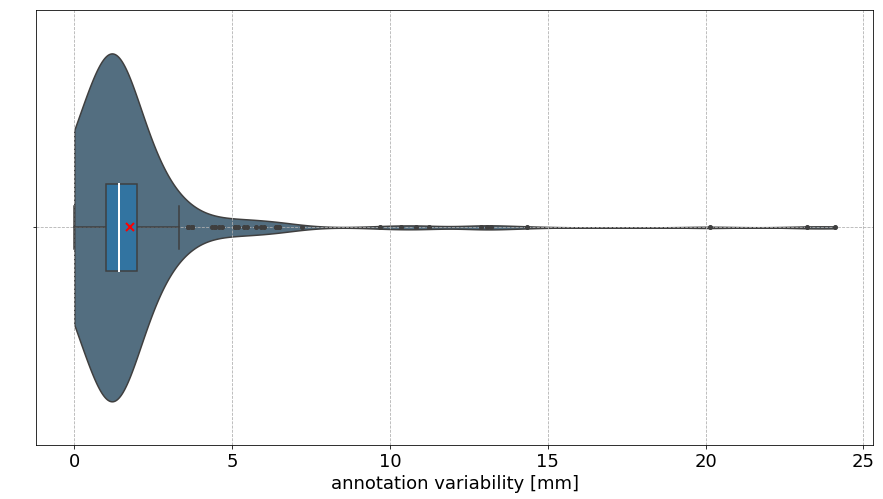}}
             \caption{Inter-rater annotation variability on a selection of test cases}
             \label{fig:IR_variability}
         \end{subfigure}
         \hspace{1mm}
         \begin{subfigure}[b]{0.49\textwidth}
             {\includegraphics[width=\textwidth]{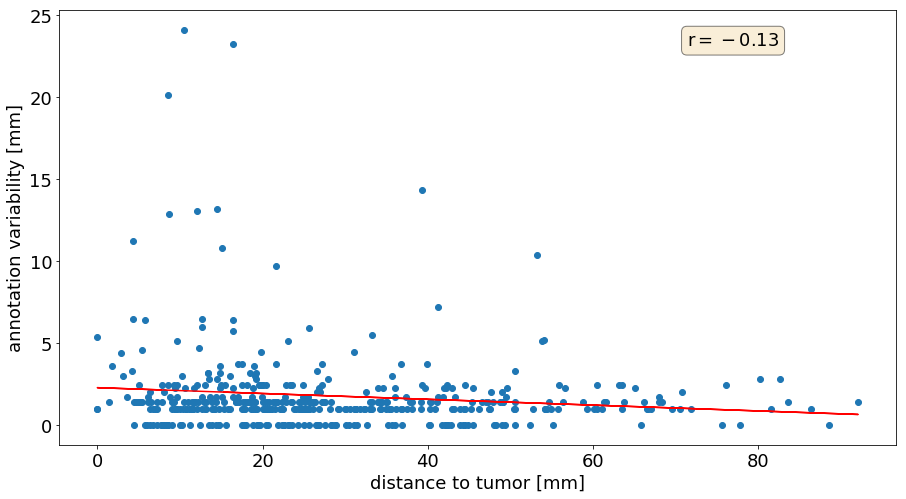}}
             \caption{Correlation of inter-rater annotation variability to the landmark-tumor distance}
             \label{fig:cor_IR_tumor}
         \end{subfigure}
         \caption{Inter-rater analysis.
         (a) shows the distribution of inter-rater annotation variability over a selection of test cases, with $median=1.41$ and $mean=1.77$.
         This variability is correlated to the respective landmark-tumor distance (blue) in (b). The respective regression line (red) and a Pearson correlation coefficient of $r=-0.13$ show a small correlation between annotation variability and landmark-tumor distance.
         }
         \label{fig:inter-rater variability}
\end{figure}
To enable the assessment of variance in our annotation process, a second expert re-annotated landmarks in follow-up scans for a subset of $21$ subjects from the dataset, leading to $446$ re-annotated landmarks in total.
The re-annotation allowed us to assess the annotation variability (AV) and provided an important baseline for comparing differences in tracking errors between various methods.
Fig. \ref{fig:IR_variability} illustrates the distribution of inter-rater variability for the selected 21 samples.
For the majority of evaluated landmarks, the AV between clinical experts lied in the range of $1-4$mm, sometimes reaching up to $24$mm.
To investigate whether landmarks in the vicinity of the tumor are more difficult to annotate, we correlated landmark-wise AV with respective distance to the tumor core (cf. Fig. \ref{fig:cor_IR_tumor}).
We observed increased levels of variability, especially in the close vicinity of the tumor. 
As the distance increased from the tumor, the variability decreased.
A Pearson correlation coefficient of $-0.13$ supported these findings, postulating a small correlation.

\subsection{Data Partitioning in Distinct Independent Cohorts}
Following the paradigm of algorithmic evaluation in the machine learning domain, the complete dataset was divided randomly and proportionally into training (70\%), validation (10\%), and testing/ranking cohorts (20\%), while taking into consideration the 1) varying deformation levels (as estimated by the GT annotations), 2) time-window between baseline and follow-up scans, and 3) number of landmarks. 

    \subsubsection{Training Data}
    The training cohort consisted of 140 cases, with each case consisting of $4$ baseline $(B)$ and $4$ follow-up $(F)$ scans, where corresponding four sequences have been acquired during the same session.
    The challenge participants were provided with the mpMRI scans and the unique landmark point locations on the baseline, as well as on the follow-up scan. 
    The participants were expected to use this data for developing and evaluating their image registration methods and subsequently submit results to the online evaluation portal.

    \subsubsection{Validation Data}
    The validation data of 20 cases was provided to the participants as scan pairs of $B$ and $F$ with landmarks provided only for the $F$ scan. Participants were requested to submit:
    \begin{itemize}
        \item Coordinates of warped landmark locations in the $B$ scan: coordinates needed to be submitted in $x,y,z$ format and had to be stored in a comma separated value (.csv) file with each row corresponding to a different landmark. 
        \item Determinant of the Jacobian of the displacement field: This needed to be submitted as a NIfTI image with the same geometry as the $B$ scan. 
    \end{itemize}

    \subsubsection{Test Data}
    The test data contained 50 cases and followed the same format and submission requirements as the validation data. 
    Additionally, a second follow-up timepoint (follow-up 2) was obtained for a subset of 9 subjects from the testing cohort to perform a longitudinal performance analysis of the submitted methods. 
    For follow-up 2, we included the same data as the other timepoints, i.e., T1, T1-CE, T2, and FLAIR scans and landmark points annotated by clinical experts.
    This data was neither released to the participants nor it is intended to be publicly released at any point.
    
    \subsubsection{Containerization of Participants' Methods}
    The participants were requested to containerize their methods using Singularity containers\footnote{\url{https://docs.sylabs.io/guides/3.5/user-guide/introduction.html}} and submit them to the challenge organizers for evaluation and ranking. 
    The organizers ran all the methods on the test data using their own computational infrastructure. This enabled confirmation of reproducibility and maximization of the use of these algorithms towards offering publicly available solutions for this problem. 
    Detailed and specific instructions for creating singularity containers of uniform Application Programming Interface (API) for this challenge were provided \footnote{\url{https://github.com/satrajitgithub/BraTS_Reg_submission_instructions}}. 
    There were some differences in the testing phase of ISBI and MICCAI as follows:
    
    \begin{itemize}
        \item ISBI BraTS-Reg Challenge\\
        The testing cohort consisted of $43$ scan pairs. Participants were asked to containerize their algorithms such that they produced i) warped landmark locations in the $B$ scan and ii) determinant of the Jacobian of displacement field.
        \item MICCAI BraTS-Reg Challenge\\
         We included seven additional heavily annotated cases (upto 50 landmarks) in the testing set leading to a total of 50 cases. 
         Also, instructions to prepare the containers were updated such that they should produce the following additional outputs along with the i) and ii) above:
         \begin{itemize}
             \item Forward and backward displacement fields.
             \item Follow-up sequences registered to baseline sequences.
             \item \verb=apply_deformation()= python function to apply a user-specified deformation field on any given input image or segmentation mask and return the output image or mask.
         \end{itemize}
         Having the additional outputs like the displacement field and the \verb=apply_deformation()= function gives flexibility in using these containers for additional analyses and applications.
    \end{itemize}

    \subsubsection{External Out of Distribution (OOD) Test Data}
    Another subset of $49$ cases was reserved to evaluate the generalizability performance of various methods on an unseen Out of Distribution (OOD) test data. Notably, this OOD dataset was not used for the ranking of the participants, in line with the original design document of the challenge \citep{https://doi.org/10.5281/zenodo.6362420}.

\subsection{Participation Policies \& Awards}
The participating teams were allowed to submit the results multiple times for the training and validation phase. 
However, for the testing phase, they were allowed to submit the container only once. 
Participants were contacted if the container resulted in errors and these errors were resolved in discussion with the participants. 
In order to be considered for ranking and award, participants were also asked to submit a short manuscript describing their methods and results. 
Containers were evaluated only if participants submitted the accompanying manuscript. 
Furthermore, the manuscript was considered for publication only if the teams successfully submitted the working container. 
Lastly, all participants were offered to be included as co-authors of this post-challenge meta-analysis manuscript.

    \subsubsection{Public Release of Participating Methods}
    Participants were asked to compete in this challenge only with fully automated algorithms. 
    User interaction (e.g., semi-automated methods) was not allowed, as the participants needed to submit their method in a singularity container to be evaluated by the organizers on the hidden testing data. 
    This was done to ensure that these algorithms can also be used in the future by anyone following the specific instructions on how to use the participants' containers. 
    By submitting their containerized algorithm, challenge participants agreed to provide their written permission to make their containers publicly available \footnote{\url{https://cloud.sylabs.io/library/search?q=brats_reg}}.

    \subsubsection{Awards}
    Participants from the organizers' research groups could participate in the challenge but were not eligible for any awards. 
    The top three performing methods from each challenge were asked to give 10-minute presentations in the main event describing their method in order to receive the award.

    \subsubsection{Use of Additional Data}
    Participants were allowed to augment the provided dataset with additional public and/or private data for scientific publication purposes. 
    However, it was mandatory for them to explicitly mention this in their submitted manuscripts. 
    Importantly, it was made clear that participants who would proceed with such an extension of the challenge dataset, must also report results using only the challenge dataset and discuss any potential differences in results in their manuscript. 
    This was due to our intentions to identify if additional data could contribute to providing an improved solution as well as to provide a fair comparison among the participating methods.


\section{Participating Methods}
In the ISBI 2022 challenge, a total of 79 teams registered, allowing them to download the data and submit the results for the training and validation phase. 
Finally, 6 teams submitted the singularity containers for the testing phase. 
In MICCAI 2022, the number of registered teams increased to 110 of which 9 teams successfully submitted their singularity containers. 
Four teams participated in both ISBI and MICCAI challenges. 

After the conclusion of the challenge, a few external groups working in the image registration domain were invited to develop methods for the BraTS-Reg challenge. 
These additional invited teams were also required to submit their singularity containers for evaluation on testing data. A request from one of the participating teams to evaluate a second version of their container after the completion of the challenge was accommodated and the method is suffixed with the term post-challenge (pc) to distinguish it from other methods. 
 Table \ref{tab:methods summary} gives an overview of the participating methods. Detailed descriptions of the evaluated methods can be found in Appendix A. Invited methods are indicated with *. 
\begin{landscape}

\begin{table}[tbp]
\scriptsize
    \caption{Summary of participating methods. 
    BCD - block coordinate descent, CR - curtavure regularizer, DF - displacement field, DR - diffusive regularization, GD - gradient descent, GE - gradient error, (G)IC - (gradient) inverse consistency, HM - histogram matching, IN - intensity normalization, IO - Instance Optimization, JL - jacobian loss, LL - landmark loss, (L)(N)CC - (local) (normalized) cross correlation, MI - mutual information, MSE - mean squared error, N/A - no information provided, NGF - normalized gradient fields, RL - reconstruction loss, SM - Symmetric Cropping, VCC - volume change control.}
      \centering
      \scriptsize
      \begin{tblr}{colspec={lcr},colspec={Q[1.0,l] Q[1.0,c] Q[1.0,c] Q[1.0,c] Q[1.0,c] Q[1.0,c] Q[1.0,c] Q[1.0,c] Q[1.0,c]}}
      \hline
        Team & MRI sequences & Loss function & Optimizer & Affine prealignment & Landmarks & Pre-processing & Post-processing & Data augmentation \\ \hline
        \tikzbullet{atomictangerine}{atomictangerine} AGHSSO & T1, T1CE, T2, FLAIR & NCC, DR, IC & Adam & Yes & No & IN & N/A & N/A \\  
        \tikzbullet{lightyellow}{lightyellow} ANTs* & T1CE & CC & GD & Yes & No & N/A & N/A & N/A \\
        \tikzbullet{lightpink}{lightpink} BDAV & T1, T1CE & NCC, MSE & Adam & No & Yes & IN & N/A & random pairing \\
        \tikzbullet{lightviolet}{lightviolet} CaMed & T1CE & GE, NCC & Adam & Yes	& Yes &	IN &	N/A & random cropping patches \\
        \tikzbullet{amaranth}{amaranth} cwmok & T1, T1CE, T2, FLAIR & NGF, LNCC & ~ & Yes & No & IN & IO & rotation, translation \\
        \tikzbullet{pureorange}{pureorange} GradICON* & T1, T1CE, T2, FLAIR & LNCC, GIC & Adam & No & No & IN & IO & translation \\
        \tikzbullet{aquamarine}{aquamarine} HyperMorph* & T1, T1CE, T2, FLAIR & NCC & Adam & Yes & No & IN & N/A & linear transforms, Gaussian noise, gamma-exponentiating  \\
        \tikzbullet{pureblue}{pureblue} Kurtlab & T1, T1CE, T2, FLAIR & MSE, DR & Adam & Yes	& No & IN, HM & Gaussian smoothing & flipping, rotation \\
        \tikzbullet{palepink}{palepink} MeDAL & T1, T1CE & NCC, MI, L2 & Adam & No & Yes & SM, IN & N/A & rotation, scaling, translation. \\
        \tikzbullet{byzantine}{byzantine} MEVIS & T1CE, T2 & NGF, CR, VCC & quasi-Newton l-BGFS & No & No & N/A & N/A & N/A \\
        \tikzbullet{lightblue}{lightblue} MICS* & T1, T1CE, T2, FLAIR & RL, NCC, JL, LL, MSE & Adam & No & Yes & N/A & N/A & N/A \\
        \tikzbullet{limegreen}{limegreen} SuperX & N/A & N/A & N/A & N/A & N/A & N/A & N/A & N/A \\
        \tikzbullet{paleblue}{paleblue} SynthMorph* & T1, T1CE, T2, FLAIR & NCC, MSE, DR & Adam & No & Yes & IN & N/A & spatial deformation, shift, rotation, scaling, shear, Gaussian noise, gamma augmentation, bias field, blurring \\
        \tikzbullet{Darkcyan}{Darkcyan} UZL & T1CE & MIND-MSE, DR & Adam & No & No & N/A & N/A & N/A \\
        \tikzbullet{aurometalsaurus}{aurometalsaurus} YKW & T1 & DF Similarity/Regularity & BCD & No & No & SM & N/A & N/A
        \end{tblr}
    \label{tab:methods summary}
\end{table}
\end{landscape}

\section{Quantitative Performance Evaluation and Ranking}
\subsection{Evaluation Metrics}
The assessment of the registration between the two scans was based on manually seeded landmarks (i.e., GT annotations) in both the pre-operative and the follow-up scans, carried out by the expert clinical neuroradiologists. 
The performance of the developed registration methods was quantitatively evaluated based on distance metrics such as Manhattan distance, Euclidean distance, and Robustness ($R$). 
While the Manhattan distance was used for creating the ranking during the ISBI 22 and MICCAI 22 challenges, the Euclidean distance was considered for detailed analysis in this manuscript, to be inline with other challenges on medical image registration \citep{borovec2020anhir,xiao2019evaluation}. 
All quantitative metrics were applied on the GT annotations in the baseline images. 
Additionally, the smoothness of the displacement field was assessed as a separate criterion.

\subsubsection{Error Distance Metrics}
For each pair $p \in P$ of baseline $(B)$ and follow-up $(F)$ scans, we refer to the landmarks in the $B$ scan as $x_l^B$ and the corresponding landmarks in the $F$ scan as $x_l^F$. 
Here, $l\in L$ where $L$ is the total number of landmarks in each scan. 
The participants were provided with $x_l^F$ and asked to estimate the coordinates of the corresponding landmark points in scan $B$ ${(\hat{x}}_l^B)$. 
The performance was then evaluated in terms of the Euclidean error (Euclidean distance). 
The Median Euclidean Error ($MEE$) between the submitted coordinates ${(\hat{x}}_l^B)$ and the manually defined coordinates (ground truth)  ${(x}_l^B)$\ was then defined as:
\begin{equation*}
MEE= Median\hspace{1mm}_{l \in L}( ||x_{l}^{B} - \hat{x}_{l}^{B}||_2)
\end{equation*}
    
\subsubsection{Robustness $(R)$}
All the participating algorithms were also evaluated according to the metric of robustness $(R)$. 
Similar to a successful-rate measure, we defined $R$ as a relative value describing how many landmarks improved their $MEE$ after registration, compared to their $MEE$ before registration. 
Let us call $K^{B, F} \subseteq L^F$ the set of successfully registered landmarks, i.e., those for which the registration error decreased. Where $L^F$ indicates the number of landmarks in the follow up scan. 
So, we define robustness $(R)$ for a pair of scans $(p)$ as the relative number of successfully registered landmarks:
\begin{equation*}
    R^{B,F}(p)=\frac{|K^{B,F}|}{|L^{F}|}
\end{equation*}
Then the average robustness over all scan pairs $(P)$ in the particular cohort is:
\begin{equation*}
    R=\frac{1}{P}\sum_{p=1}^{P}R^{B,F}(p)
\end{equation*}
Therefore, $R$ is the relative value (in the range of $0$ and $1$) of how many landmarks across all scan pairs have an improved $MEE$ after registration, when compared to the initial $MEE$. 
When $R$ is equal to $1$, the average distance of all the landmarks in the target and warped images is reduced after registration, whereas $0$ means that none of the distances is reduced.

\subsubsection{Smoothness of the Displacement Field}
We also evaluated the smoothness of the displacement field by calculating its  determinant of the Jacobian and examining the number and percentage of voxels with a non-positive Jacobian determinant for each method. 
These voxels correspond to locations where the deformation is not diffeomorphic. 
Furthermore, we computed both the minimum and the $99^{th}$ percentile of the Jacobian determinant (as opposed to the maximum, which is susceptible to noise).
These were computed within different regions of interest (e.g., within the tumor core, within 30mm from the tumor core boundary, outside 30mm from the tumor core boundary, etc.). 
This metric was not planned to be used to determine the ranking of the participating algorithms during the challenge. 
Instead, in the event of a tie between teams, the one with the higher smoothness of displacement field would be ranked higher.

\subsubsection{Baseline Performance}
The baseline performance for registration between the given pre-operative baseline and follow-up scans was established using an affine registration. 
This decision was driven by the comparative analysis of various methods shown in \citep{han2018patient}, where every deformable registration method performed better than affine registration but worse than human experts.
The baseline registration was performed using the default parameters of an affine ANTs registration (\href{https://github.com/ANTsX/ANTs/blob/master/Scripts/antsRegistrationSyN.sh}{affine baseline}). 
Participating methods that performed worse than the affine registration results were considered for the BraTS-Reg challenge, but were acknowledged to state this clearly in their manuscript.

\subsection{Patient-wise Ranking Strategy and the BraTS-Reg score}
\label{bratsreg_score}
We observed variations in the ranking of different teams based on whether the $MEE$ or Robustness metric was taken into account. 
Hence, we introduced a new ranking score termed the ``BraTS-Reg score", which consolidates individual rankings. 
In this process, each participating team was ranked across the two evaluation metrics ($MEE$ and $R$) and for each case from the entire testing set (comprising $50$ cases for MICCAI), resulting in $100$ rankings per team. 
The ultimate ranking was determined by the average of these rankings, normalized by the number of teams, and was referred to as the BraTS-Reg score.

\subsection{Longitudinal Analysis} 
\label{longitudinal_description}
The goal of longitudinal analysis was to check the performance robustness of the proposed methods across different timepoints.
For this purpose, the following four different registration tasks were performed (where $m$ to $f$ signifies $m$ as the moving scan and $f$ as the fixed target scan):
\begin{enumerate}
    \item Follow-up 1 to baseline (standard task)
    \item Follow-up 2 to baseline
    \item Follow-up 2 to Follow-up 1
    \item Follow-up 2 to Follow-up 1 to baseline
\end{enumerate}

Tasks 1, 2 and 3 could be performed in a single operation whereas task 4 was performed in two stages. 
In the first stage, the follow-up 2 scans were warped to follow-up 1, and subsequently, in the second stage, these scans were further warped to the baseline scans. 
Task 4 aimed to identify the potential for error propagation in registration across multiple time points. 
To that end, the registration performances of the proposed methods were analyzed for the four tasks and the consistency of the different evaluation metrics was compared across these tasks to perform additional meta-analysis after the conclusion of the challenge. 
Similar to the analysis performed for the ranking evaluation, the longitudinal performances were assessed in terms of MEE, Robustness, and the BraTS-Reg score.  

\subsection{Analysis on the Relation of Performance and Inter-Rater Annotation Variability}\label{IR_methods_sec}
For further analysis with respect to the inter-rater annotation variability as described in Section \ref{IR_sec}, we performed two additional analyses, similar to \citep{waldmannstetter2023framing}. 
In the first analysis, we defined a spherical region of interest (ROI) for each reference ground truth landmark $({x}_l^B)$ in a selected subset from the dataset that was prepared for inter-rater analysis, including $21$ cases with $446$ re-annotated landmarks from the testing cohort. 
Each ROI was sized depending on its respective annotation variability ${AV}_l^B$. 
This variability was therefore defined as the Euclidean error (EE) between the annotations of two raters $({x}_{r1}^B)$ and $({x}_{r2}^B)$ for the same landmark:
\begin{equation*}
    {AV}_l^B = \text{ROI $({x}_l^B) = EE_{AV}({x}_{r1}^B, {x}_{r2}^B)$}
\end{equation*}
with $EE_{AV}({x}_{r1}^B, {x}_{r2}^B) = ||{x}_{r1}^B - {x}_{r2}^B||_2$. 
We then denoted a \emph{hit} as the event (success), when the EE between the reference landmark $({x}_l^B)$ and the warped landmark $(\hat{x}_l^B)$ (produced by a participating algorithm) was lower than the respective reference landmark's annotation variability:
\begin{equation*}
    {(\hat{x}}_l^B)=
    \begin{cases}
    hit, & \text{if $(\hat{x}_l^B)\in$ ROI $({x}_l^B)$}\\
    miss, & \text{if $(\hat{x}_l^B)\notin$ ROI $({x}_l^B)$}\\
    \end{cases}
\end{equation*}
Then, we computed the ratio of \emph{hits} and \emph{misses}, giving us the \emph{hit rate} of the respective algorithm. 
As already mentioned above, MEE and Robustness were initially assessed on the ground truth landmarks in the baseline images. 
For the inter-rater analysis, follow-up images were used. Kindly note that due to the difference in image sources, the results may not be directly comparable. 

In a second analysis, we made use of the distribution ($D$) of the inter-rater annotation variability, as described in section \ref{IR_sec} and visualized in Fig. \ref{fig:IR_variability}.
We therefore computed \emph{hit rate} curves \citep{waldmannstetter2023framing}, by sampling thresholds from $D$ using the formula:
\begin{equation*}
    median (D) + \delta \cdot median\ absolute\ deviation (D)
\end{equation*}
with $\delta \in [\delta_{min}, \delta_{max}]$ for positive thresholds. 
$[\delta_{min}, \delta_{max}]$ was therefore defined with respect to the desired evaluation range. 
Here, $\delta$ was chosen to be in $[-1, 10]$ using steps of $0.5$, with $median=1.41mm$ and $median\ absolute\ deviation=0.77mm$.
The goal of this evaluation was to compare the performance of the automated algorithms against the annotation variability of human experts, representing the ``gold standard'' in such tasks.

\section{Results}
\subsection{Combined Results of ISBI + MICCAI Challenge}
\begin{figure}[!t]
         \centering
         \begin{subfigure}[b]{0.49\textwidth}
             {\includegraphics[width=\textwidth]{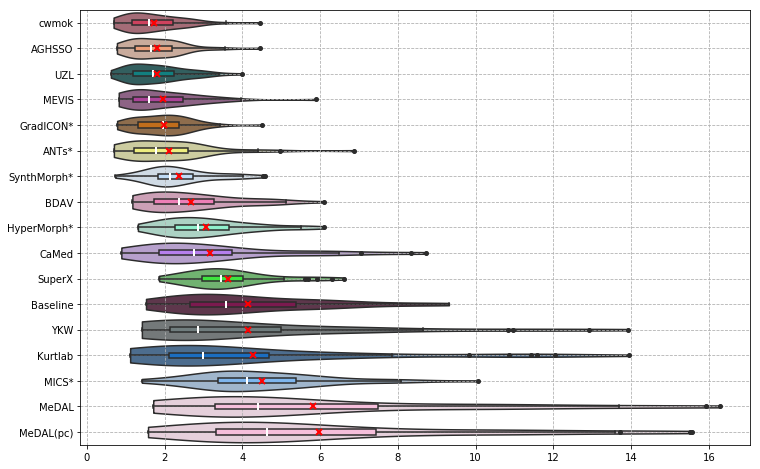}}
             \caption{Performance comparison based on MEE on test data}
             \label{fig:test set MEE}
         \end{subfigure}
         \hspace{1mm}
         \begin{subfigure}[b]{0.49\textwidth}
             {\includegraphics[width=\textwidth]{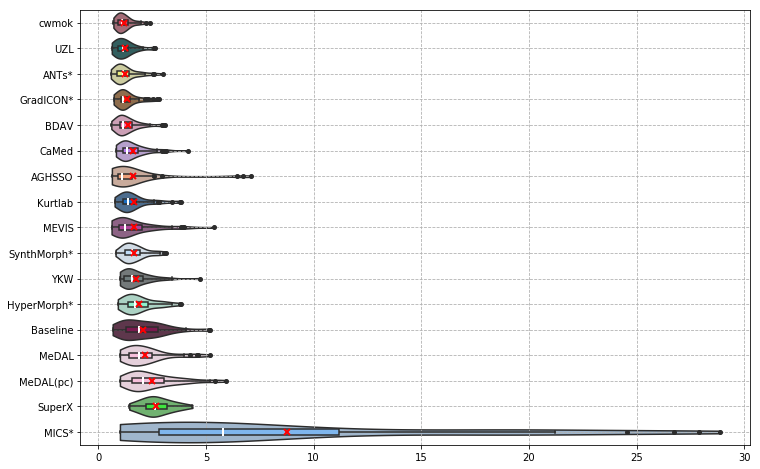}}
             \caption{Performance comparison based on MEE on OOD data}
             \label{fig:OOD set MEE}
         \end{subfigure}
         \\
         \begin{subfigure}[b]{0.49\textwidth}
             {\includegraphics[width=\textwidth]{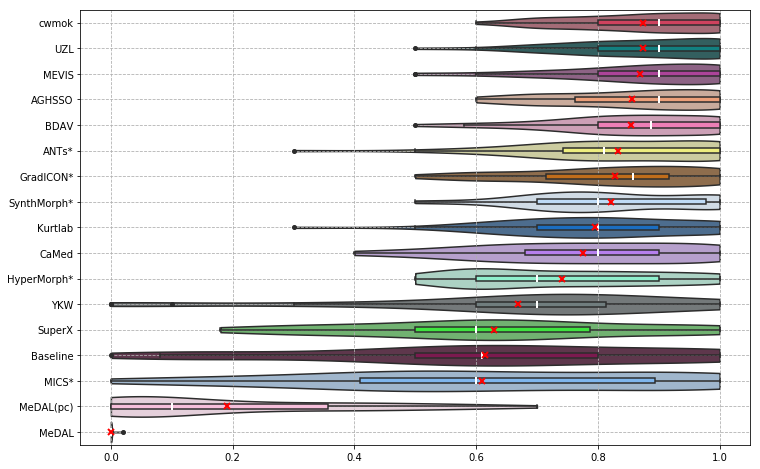}}
             \caption{Performance comparison based on Robustness on test data}
             \label{fig:test set R}
         \end{subfigure}
         \hspace{1mm}
         \begin{subfigure}[b]{0.49\textwidth}
             {\includegraphics[width=\textwidth]{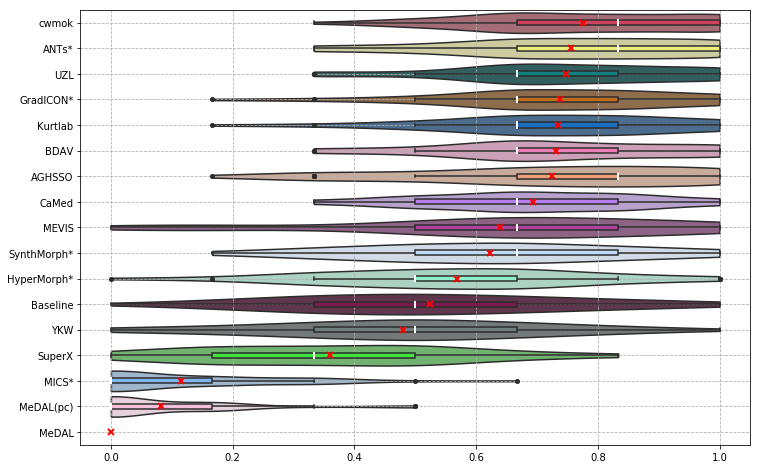}}
             \caption{Performance comparison based on Robustness on OOD data}
             \label{fig:OOD set R}
         \end{subfigure}
         \\
         \begin{subfigure}[b]{0.49\textwidth}
             {\includegraphics[width=\textwidth]{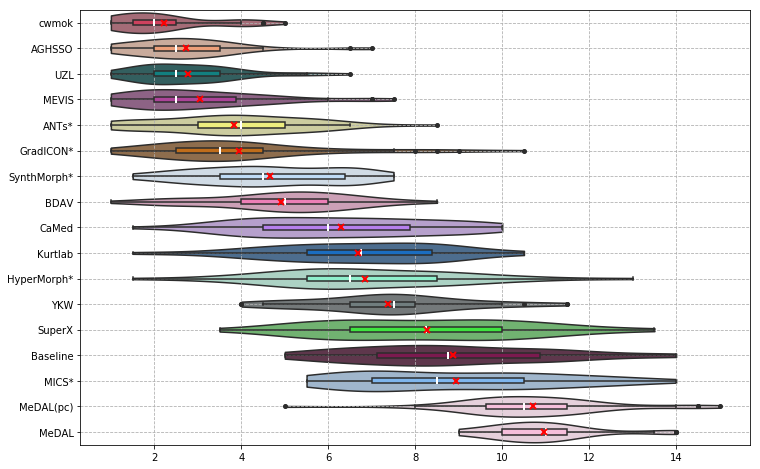}}
             \caption{Performance comparison based on BraTS-Reg score on test data }
             \label{fig:test set BraTSReg score}
         \end{subfigure}
         \hspace{1mm}
         \begin{subfigure}[b]{0.49\textwidth}
             {\includegraphics[width=\textwidth]{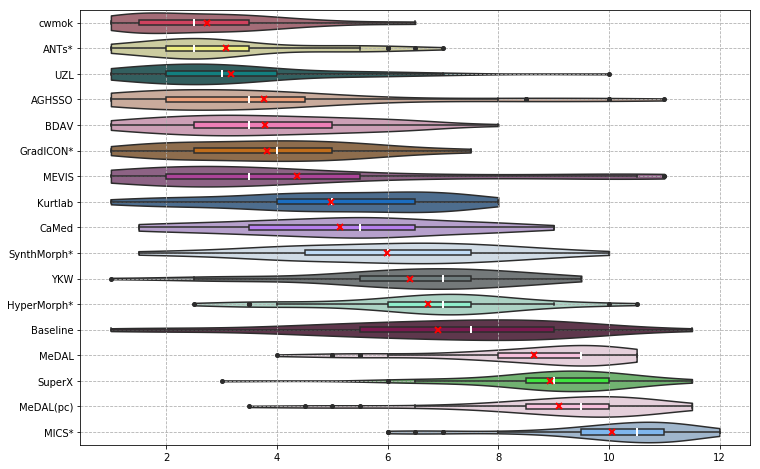}}
             \caption{Performance comparison based on BraTS-Reg score on OOD data}
             \label{fig:OOD set BraTSReg score}
         \end{subfigure}
         
         \caption{Comparative performance analysis of various participating methods in terms of Median Euclidean Error (MEE), Robustness and BraTS-Reg score  along with the invited methods (indicated with *). The white line in the violin plots indicates the median, whereas the red cross indicates the mean  of the distribution.}
         \label{fig:analysis on test data}
\end{figure} 
\begin{table}[!t]
      \caption{Overview of a variety of evaluation metrics per team on the test set. Euclidean Errors (EE) are given in mm, while Robustness is given in range $[0,1]$.}
      \centering
      \scriptsize
      \begin{tblr}{colspec={lcr},colspec={Q[1.0,c] Q[1.0,c] Q[1.0,c] Q[1.0,c] Q[1.0,c] Q[1.0,c] Q[1.0,c]}}
      \hline
        ~ & Median of Median EE & Mean of Median EE & Median of Mean EE & Mean of Mean EE & Median Robustness & Mean Robustness \\ \hline
        AGHSSO & $1.66$ & $1.79$ & $2.26$ & $2.51$ & $0.90$ & $0.86$ \\
        ANTs\textsuperscript{*} & $1.78$ & $2.10$ & $2.75$ & $2.96$ & $0.81$ & $0.83$ \\
        BDAV & $2.38$ & $2.68$ & $3.13$ & $3.43$ & $0.89$ & $0.85$ \\
        CaMed & $2.76$ & $3.17$ & $3.47$ & $4.06$ & $0.80$ & $0.78$ \\
        cwmok & $1.60$ & $1.73$ & $2.04$ & $2.38$ & $0.90$ & $0.87$ \\
        MEVIS & $1.59$ & $1.95$ & $2.44$ & $2.75$ & $0.90$ & $0.87$ \\
        GradICON\textsuperscript{*} & $1.95$ & $1.97$ & $2.61$ & $2.73$ & $0.86$ & $0.83$ \\
        HyperMorph\textsuperscript{*} & $2.85$ & $3.072$ & $3.60$ & $3.81$ & $0.70$ & $0.74$ \\
        Kurtlab & $2.99$ & $4.26$ & $3.87$ & $5.07$ & $0.80$ & $0.79$ \\
        MeDAL & $4.39$ & $5.81$ & $5.21$ & $6.37$ & $0.00$ & $0.00$ \\
        MeDAL(pc) & $4.62$ & $5.96$ & $5.32$ & $6.47$ & $0.10$ & $0.19$ \\
        MICS\textsuperscript{*} & $4.12$ & $4.51$ & $4.75$ & $5.15$ & $0.60$ & $0.61$ \\
        SuperX & $3.46$ & $3.62$ & $4.10$ & $4.39$ & $0.60$ & $0.63$ \\
        SynthMorph\textsuperscript{*} & $2.15$ & $2.36$ & $2.81$ & $3.06$ & $0.80$ & $0.82$ \\
        UZL & $1.71$ & $1.79$ & $2.36$ & $2.42$ & $0.90$ & $0.87$ \\
        YKW & $2.86$ & $4.15$ & $3.68$ & $4.81$ & $0.70$ & $0.67$ \\ 
        Baseline & $3.59$ & $4.14$ & $4.48$ & $4.86$ & $0.61$ & $0.61$ \\
        \hline 
     \end{tblr}
     \label{tab:summary_results}
     \end{table} 
Comparative performance analysis of various participating methods in terms of MEE, Robustness, and BraTS-Reg score in ISBI and MICCAI 2022 along with the invited methods are shown separately on the actual test data and out-of-distribution (OOD) test data in Fig. \ref{fig:analysis on test data}. In all violin plots, the teams are arranged from top to bottom based on their performance, with the order reflecting the spectrum from best to worst in terms of the mean value of the corresponding evaluation metrics. 
As it can be observed from Fig. \ref{fig:test set MEE} and Fig. \ref{fig:test set R}, the team rankings are not uniform in terms of MEE and R. Team \emph{cwmok} performed best in terms of both MEE and R, clearly indicating the top-ranked team. 
However, \emph{AGHSSO}, \emph{UZL} and \emph{MEVIS} secured ranks 2, 3, and 4 in terms of MEE, while ranks varied in terms of robustness. This made it difficult to finalize the winners. 
To overcome this issue, the final ranking was based on the BraTS-Reg score generated by the patient-wise ranking strategy as described in section \ref{bratsreg_score} and is shown in Fig. \ref{fig:test set BraTSReg score}. 
Fig. \ref{fig:analysis on test data} also shows the results on out-of-distribution (OOD) data. 
A detailed overview of different team-wise evaluation metrics is given in Table \ref{tab:summary_results} along with the baseline performance obtained with affine registration using default parameters of \emph{ANTs}. 

\subsection{Statistical Analysis of the Results} 

\begin{table}[h]
    \caption{Statistical significance analysis based on p-value calculation on the test data, showing different tiers between the participating teams, demarcated by horizontal lines.}
      \centering
      \tiny
      \begin{tblr}{colspec={lcr},colspec={Q[1.0,c] Q[1.0,c] Q[1.0,c] Q[1.0,c] Q[1.0,c] Q[1.0,c] Q[1.0,c] Q[1.0,c] Q[1.0,c] Q[1.0,c] Q[1.0,c] Q[1.0,c] Q[1.0,c] Q[1.0,c] Q[1.0,c] Q[1.0,c]}}
      \hline
        ~ & cwmok & AGHSSO & UZL & MEVIS & ANTs\textsuperscript{*} & \begin{tabular}[c]{@{}c@{}}Grad\\ICON\textsuperscript{*}\end{tabular} & \begin{tabular}[c]{@{}l@{}}Synth\\Morph\textsuperscript{*}\end{tabular} & BDAV & CaMed & Kurtlab & \begin{tabular}[c]{@{}l@{}}Hyper\\Morph\textsuperscript{*}\end{tabular} & YKW & SuperX & Baseline & MICS\textsuperscript{*} & \begin{tabular}[c]{@{}c@{}}MeDAL\\ (pc)\end{tabular} & MeDAL \\
        \hline
        cwmok & - & 0.3385 & 0.1924 & 
        0.0479 & 0.0051 & 0.0144 & 0 & 0.0007 & 0 & 0.0003 & 0 & 0 & 0 & 0 & 0 & 0 & 0 \\
        AGHSSO & - & - & 0.2945 & 0.1648 & 0.0099 & 0.0299 & 0.0018 & 0 & 0 & 0.0004 & 0.0002 & 0 & 0 & 0 & 0 & 0 & 0 \\
        UZL & - & - & - & 0.2375 & 0.0113 & 0.0293 & 0.001 & 0.0003 & 0.0001 & 0 & 0.0003 & 0 & 0 & 0 & 0 & 0 & 0 \\ \hline
        MEVIS & - & - & - & - & 0.0716 & 0.2936 & 0.0299 & 0.0199 & 0 & 0.0014 & 0.0002 & 0 & 0 & 0 & 0 & 0 & 0 \\
        ANTs\textsuperscript{*} & - & - & - & - & - & 0.6445 & 0.1697 & 0.0868 & 0.0002 & 0.0153 & 0.0069 & 0 & 0 & 0 & 0 & 0 & 0 \\
        \begin{tabular}[c]{@{}c@{}}Grad\\ICON\textsuperscript{*}\end{tabular} & - & - & - & - & - & - & 0.0556 & 0.0424 & 0.0003 & 0.0096 & 0.003 & 0.0003 & 0 & 0 & 0 & 0 & 0 \\ \hline
        \begin{tabular}[c]{@{}l@{}}Synth\\Morph\textsuperscript{*}\end{tabular} & - & - & - & - & - & - & - & 0.383 & 0.0005 & 0.0329 & 0.0067 & 0.0003 & 0 & 0 & 0 & 0 & 0 \\
        BDAV & - & - & - & - & - & - & - & - & 0.0045 & 0.056 & 0.0246 & 0.0003 & 0 & 0 & 0 & 0 & 0 \\ \hline  
        CaMed & - & - & - & - & - & - & - & - & - & 0.8653 & 0.9335 & 0.276 & 0.0593 & 0 & 0.0141 & 0.0009 & 0 \\
        Kurtlab & - & - & - & - & - & - & - & - & - & - & 0.4164 & 0.0436 & 0.0148 & 0.0005 & 0.0032 & 0 & 0 \\
        \begin{tabular}[c]{@{}l@{}}Hyper\\Morph\textsuperscript{*}\end{tabular} & - & - & - & - & - & - & - & - & - & - & - & 0.0311 & 0.0022 & 0 & 0.0003 & 0 & 0 \\ \hline
        YKW & - & - & - & - & - & - & - & - & - & - & - & - & 0.2106 & 0.0045 & 0.0651 & 0.0019 & 0 \\
        SuperX & - & - & - & - & - & - & - & - & - & - & - & - & - & 0.0194 & 0.1103 & 0.0064 & 0.0003 \\ \hline
        Baseline & - & - & - & - & - & - & - & - & - & - & - & - & - & - & 0.8037 & 0.0722 & 0.0005 \\
        MICS\textsuperscript{*} & - & - & - & - & - & - & - & - & - & - &  -& - & - & - & - & 0.0089 & 0.0002 \\ \hline
        \begin{tabular}[c]{@{}c@{}}MeDAL\\ (pc)\end{tabular} & - & - & - & - & - & - & - & - & - & - & - & - & - & - & - & - & 0.0423 \\ \hline
        MeDAL & - & - & - & - & - & - & - & - & - & - & - & - & - & - & - & - & - \\ \hline
    \end{tblr}
    \label{tab:p-value Test}
\end{table}

\begin{table}[!h]
    \caption{Statistical significance analysis based on p-value calculation on the OOD data, showing different tiers between the participating teams, demarcated by horizontal lines.}
      \centering
      \tiny
      \begin{tblr}{colspec={lcr},colspec={Q[1.0,c] Q[1.0,c] Q[1.0,c] Q[1.0,c] Q[1.0,c] Q[1.0,c] Q[1.0,c] Q[1.0,c] Q[1.0,c] Q[1.0,c] Q[1.0,c] Q[1.0,c] Q[1.0,c] Q[1.0,c] Q[1.0,c] Q[1.0,c]}}
      \hline
        ~ & cwmok & ANTs\textsuperscript{*} & UZL & AGHSSO & BDAV & \begin{tabular}[c]{@{}c@{}}Grad\\ICON\textsuperscript{*}\end{tabular} & MEVIS & Kurtlab & CaMed & \begin{tabular}[c]{@{}l@{}}Synth\\Morph\textsuperscript{*}\end{tabular} & YKW & \begin{tabular}[c]{@{}l@{}}Hyper\\Morph\textsuperscript{*}\end{tabular} & Baseline & MeDAL & SuperX & \begin{tabular}[c]{@{}c@{}}MeDAL\\ (pc)\end{tabular} & MICS\textsuperscript{*} \\ \hline
        cwmok & - & 0.4913 & 0.6091 & 0.1707 & 0.0545 & 0.0398 & 0.0014 & 0.0156 & 0.0352 & 0 & 0.0001 & 0 & 0.0008 & 0 & 0 & 0 & 0 \\
        ANTs\textsuperscript{*} & - & - & 0.5973 & 0.1447 & 0.0215 & 0.0364 & 0.0014 & 0.0081 & 0.0227 & 0.0009 & 0.0004 & 0.0002 & 0.0005 & 0 & 0 & 0 & 0 \\
        UZL & - & - & - & 0.1494 & 0.0108 & 0.0186 & 0.0002 & 0.0032 & 0.0077 & 0 & 0.0001 & 0 & 0 & 0 & 0 & 0 & 0 \\
        AGHSSO & - & - & - & - & 0.2688 & 0.4315 & 0.0189 & 0.1476 & 0.1993 & 0.0077 & 0.0019 & 0.0005 & 0.0025 & 0 & 0 & 0 & 0 \\ \hline
        BDAV & - & - & - & - & - & 0.7444 & 0.0195 & 0.1983 & 0.365 & 0.0065 & 0.0025 & 0.0003 & 0.0009 & 0 & 0 & 0 & 0 \\ 
        \begin{tabular}[c]{@{}c@{}}Grad\\ICON\textsuperscript{*}\end{tabular} & - & - & - & - & - & - & 0.0094 & 0.1157 & 0.2037 & 0.001 & 0.0006 & 0 & 0.0021 & 0 & 0 & 0 & 0 \\ \hline 
        MEVIS & - & - & - & - & - & - & - & 0.959 & 0.94 & 0.5991 & 0.3582 & 0.2576 & 0.3339 & 0.015 & 0.0087 & 0.0076 & 0.0031 \\
        Kurtlab & - & - & - & - & - & - & - & - & 0.6073 & 0.0239 & 0.0016 & 0.0014 & 0.0031 & 0 & 0 & 0 & 0 \\
        CaMed & - & - & - & - & - & - & - & - & - & 0.0234 & 0.002 & 0 & 0.0002 & 0 & 0 & 0 & 0 \\ \hline
        \begin{tabular}[c]{@{}l@{}}Synth\\Morph\textsuperscript{*}\end{tabular} & - & - & - & - & - & - & - & - & - & - & 0.1826 & 0.1081 & 0.1335 & 0.0002 & 0.0001 & 0 & 0.0001 \\
        YKW & - & - & - & - & - & - & - & - & - & - & - & 0.3261 & 0.4254 & 0.0007 & 0.0001 & 0.0012 & 0.0001 \\
        \begin{tabular}[c]{@{}l@{}}Hyper\\Morph\textsuperscript{*}\end{tabular} & - & - & - & - & - & - & - & - & - & - & - & - & 0.5745 & 0.0104 & 0.0005 & 0.0034 & 0.0002 \\
        Baseline & - & - & - & - & - & - & - & - & - & - & - & - & - & 0.0036 & 0.0004 & 0.0011 & 0.0001 \\ \hline
        MeDAL & - & - & - & - & - & - & - & - & - & - & - & - & - & - & 0.1642 & 0.0152 & 0.0209 \\
        SuperX & - & - & - & - & - & - & - & - & - & - & - & - & - & - & - & 0.4743 & 0.0223 \\ \hline 
        \begin{tabular}[c]{@{}c@{}}MeDAL\\ (pc)\end{tabular} & - & - & - & - & - & - & - & - & - & - & - & - & - & - & - & - & 0.0775 \\
        MICS\textsuperscript{*} & - & - & - & - & - & - & - & - & - & - & - & - & - & - & - & - & - \\ \hline
    \end{tblr}
    \label{tab:p-value OOD}
\end{table}

Summary of results as shown in Fig. \ref{fig:analysis on test data} and Table \ref{tab:summary_results} indicates that the performance of some methods was very close to each other. 
To more accurately quantify differences among the various methods, we employed a statistical analysis procedure akin to those conducted in other challenges like Ischemic Lesion Segmentation (ISLES) \citep{maier2017isles, winzeck2018isles} and  BraTS \citep{menze2014multimodal,bakas2017advancing,bakas2018identifying,bakas2017segmentationGBM,bakas2017segmentationLGG,baid2021rsna}. 
Specifically, we performed a pairwise comparison for significant differences in the relative rankings based on $100,000$ permutations to quantify the statistical similarities among the results of different methods.
For each team, we started with the list of observed subject-level cumulative ranks, i.e., the actual ranking based on the BraTS-Reg score. For each pair of teams, we repeatedly randomly permuted (100,000 times) the cumulative ranks for each subject. 
For each permutation, we calculated the difference in the BraTS-Reg score between this pair of teams. 
The statistical significance of the relative rankings was determined by assessing the proportion of occasions when the difference in BraTS-Reg score, calculated using randomly permuted data, surpassed the observed difference in BraTS-Reg score (i.e., using the actual data). 
This proportion was reported as a p-value. 
These values are reported in an upper triangular matrix in Tables \ref{tab:p-value Test} and \ref{tab:p-value OOD} for the test and OOD data, respectively. 
The statistical evaluation of the top-ranked teams on test data revealed that the first three teams were statistically similar to each other (p $>$ 0.05). These teams were statistically better than the fourth team (p=0.0479) indicated by a horizontal line in Table \ref{tab:p-value Test}. 
Similarly, all the participating methods were divided into different tiers and separated by horizontal lines for the test and OOD data in Tables \ref{tab:p-value Test} and \ref{tab:p-value OOD}, respectively. 


\subsection{Longitudinal Analysis Results}
\begin{figure}[t]
         \centering
         \begin{subfigure}[b]{0.49\textwidth}
             {\includegraphics[width=\textwidth]{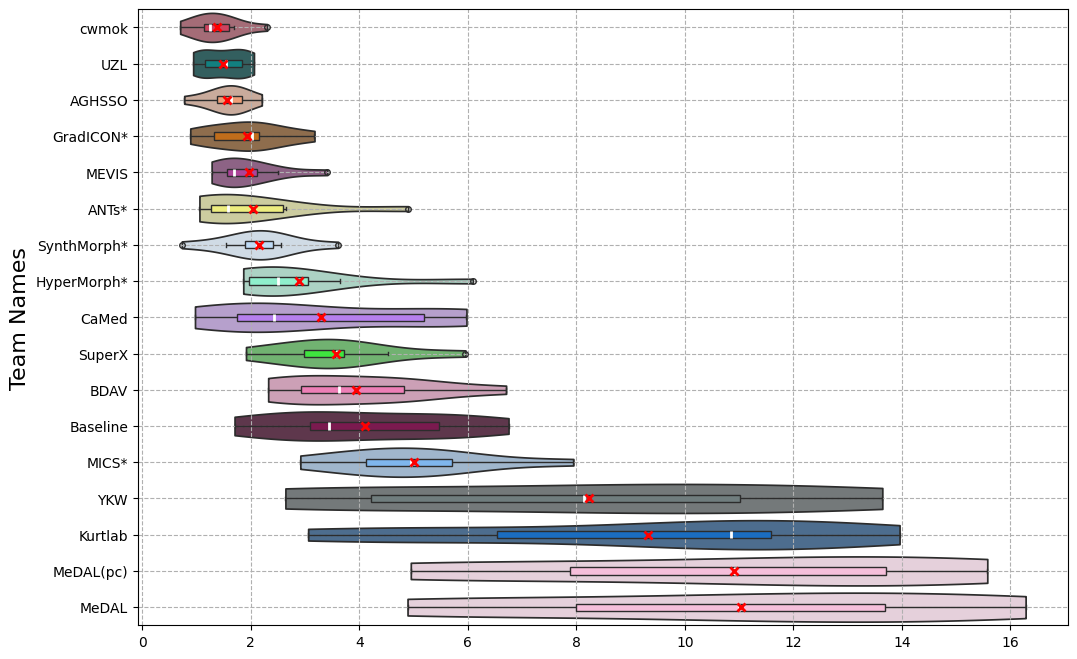}}
             \caption{MEE for follow-up 1 to baseline registration task}
             \label{fig:f1 to b MEE}
         \end{subfigure}
         \hspace{1mm}
         \begin{subfigure}[b]{0.49\textwidth}
             {\includegraphics[width=\textwidth]{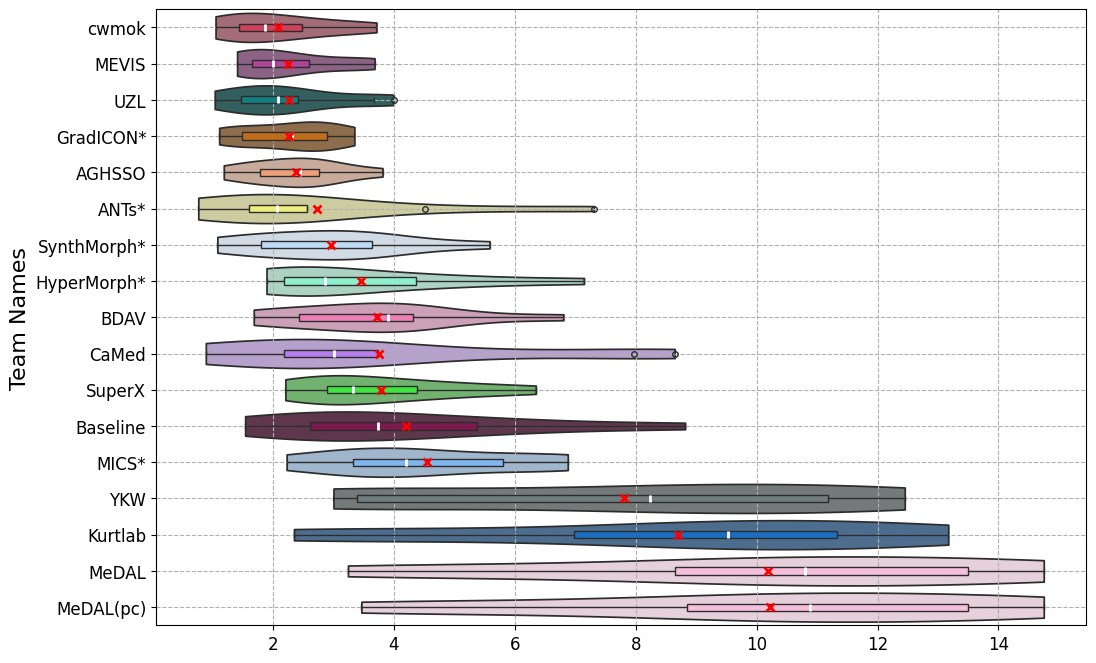}}
             \caption{MEE for follow-up 2 to baseline registration task}
             \label{fig:f2 to b MEE}
         \end{subfigure}
         \hspace{1mm}
         \begin{subfigure}[b]{0.49\textwidth}
             {\includegraphics[width=\textwidth]{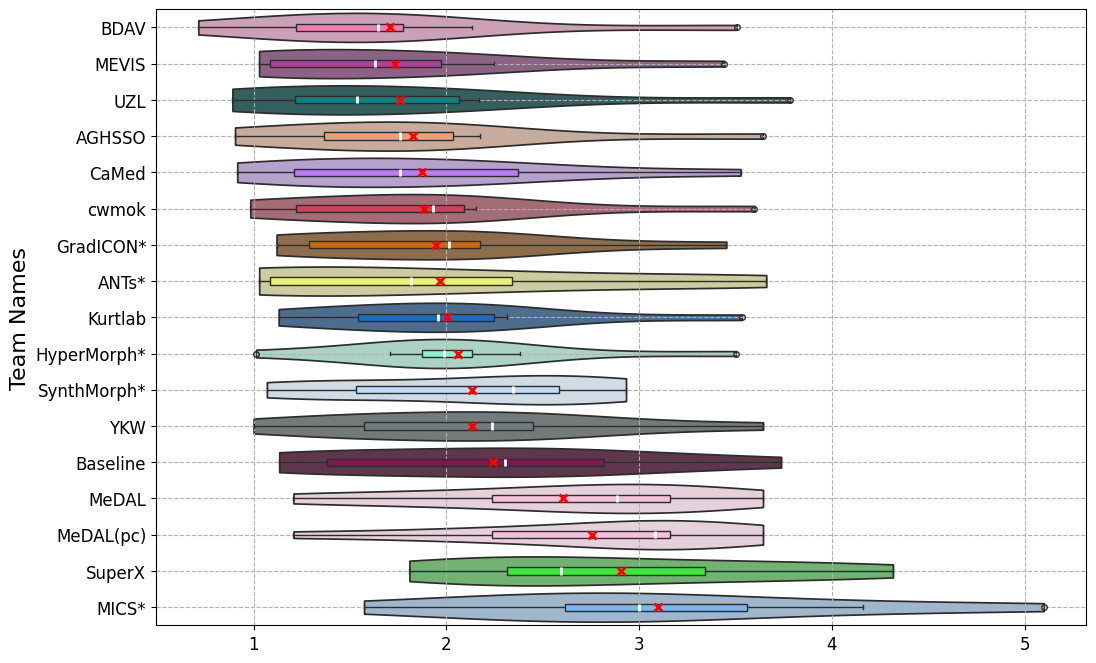}}
             \caption{MEE for follow-up 2 to follow-up 1 registration task}
             \label{fig:f2 to f1 MEE}
         \end{subfigure}
         \hspace{1mm}
         \begin{subfigure}[b]{0.49\textwidth}
             {\includegraphics[width=\textwidth]{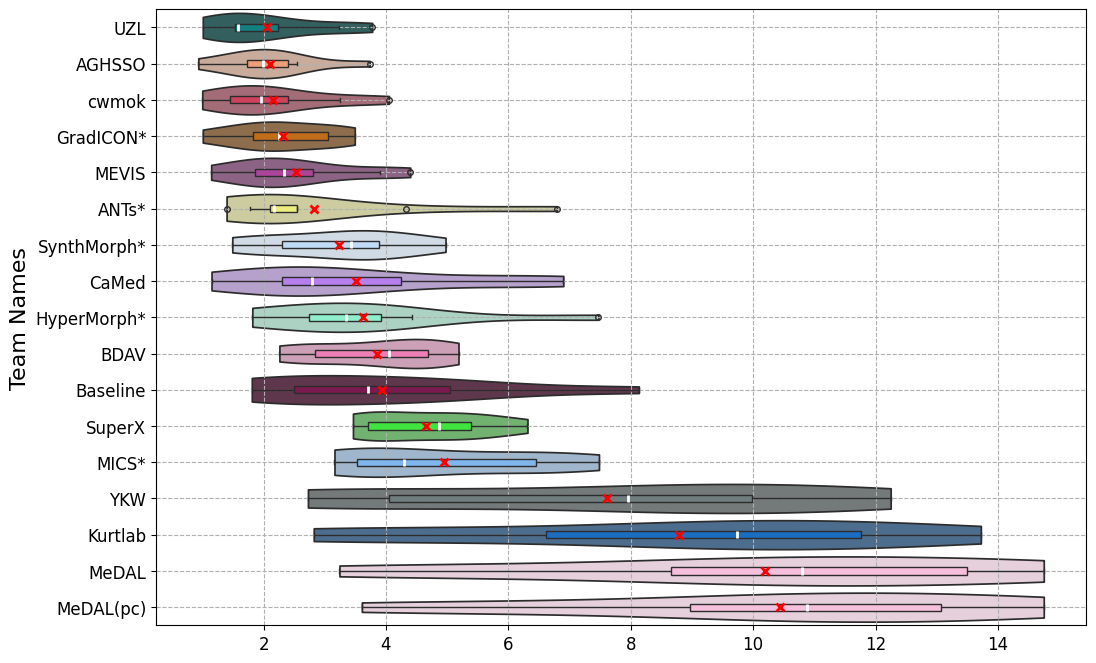}}
             \caption{MEE for follow-up 2 to follow-up 1 to baseline registration task}
             \label{fig:f2 to f1 to b MAE}
         \end{subfigure}
         \caption{Comparative performance analysis of various participating methods in terms of MEE on longitudinal data.}
         \label{fig:longitudinal analysis MAE}
      \end{figure}
      
      \begin{figure}[h]
         \centering
         \begin{subfigure}[b]{0.49\textwidth}
             {\includegraphics[width=\textwidth]{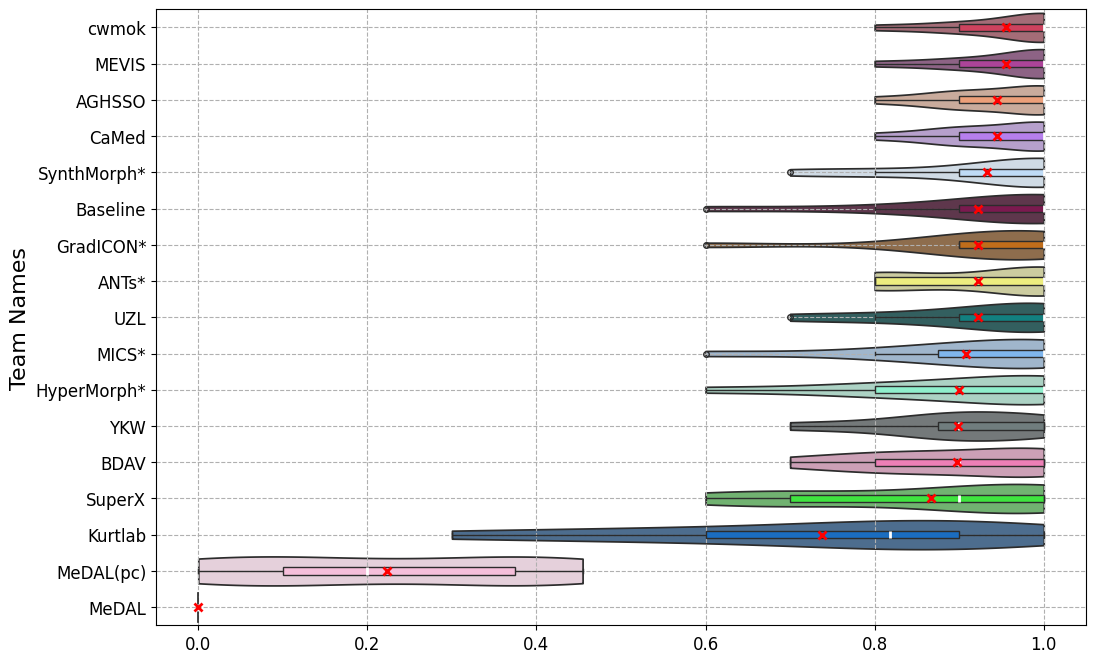}}
             \caption{Robustness for follow-up 1 to baseline registration task}
             \label{fig:f1 to b R}
         \end{subfigure}
         \hspace{1mm}
         \begin{subfigure}[b]{0.49\textwidth}
             {\includegraphics[width=\textwidth]{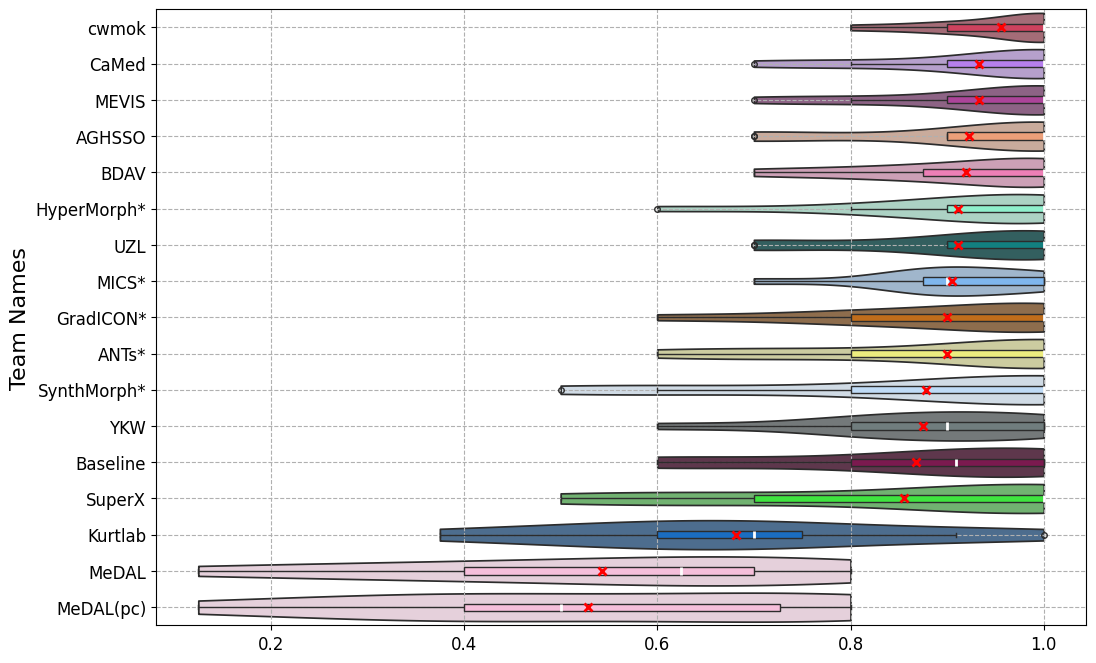}}
             \caption{Robustness for follow-up 2 to baseline registration task}
             \label{fig:f2 to b R}
         \end{subfigure}
         \hspace{1mm}
         \begin{subfigure}[b]{0.49\textwidth}
             {\includegraphics[width=\textwidth]{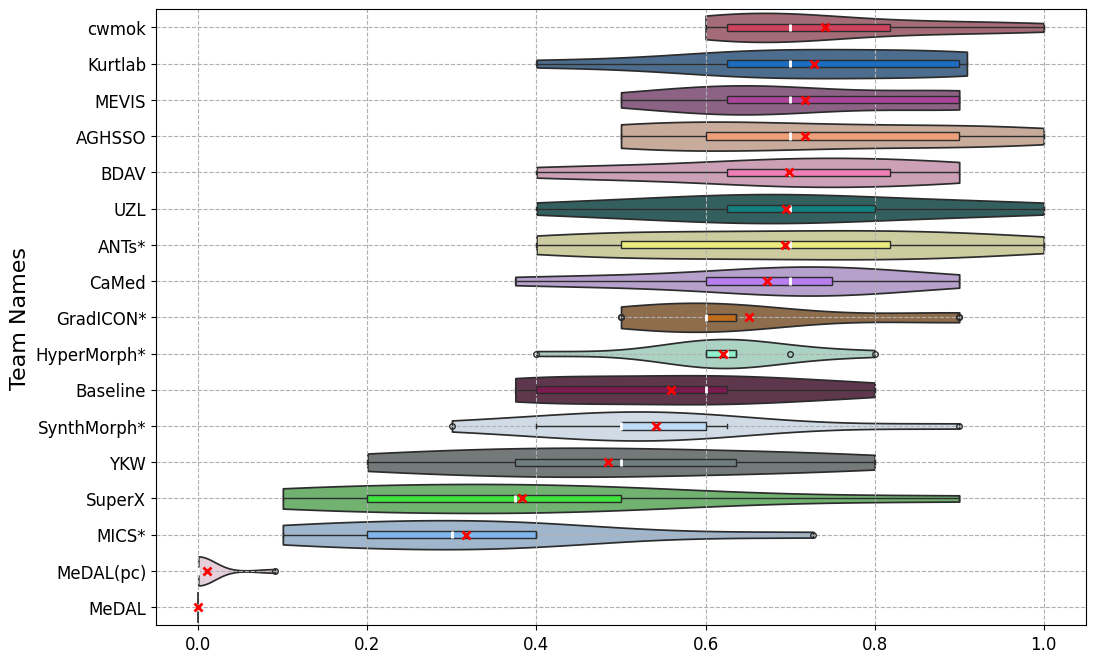}}
             \caption{Robustness for follow-up 2 to follow-up 1 registration task}
             \label{fig:f2 to f1 R}
         \end{subfigure}
         \hspace{1mm}
         \begin{subfigure}[b]{0.49\textwidth}
             {\includegraphics[width=\textwidth]{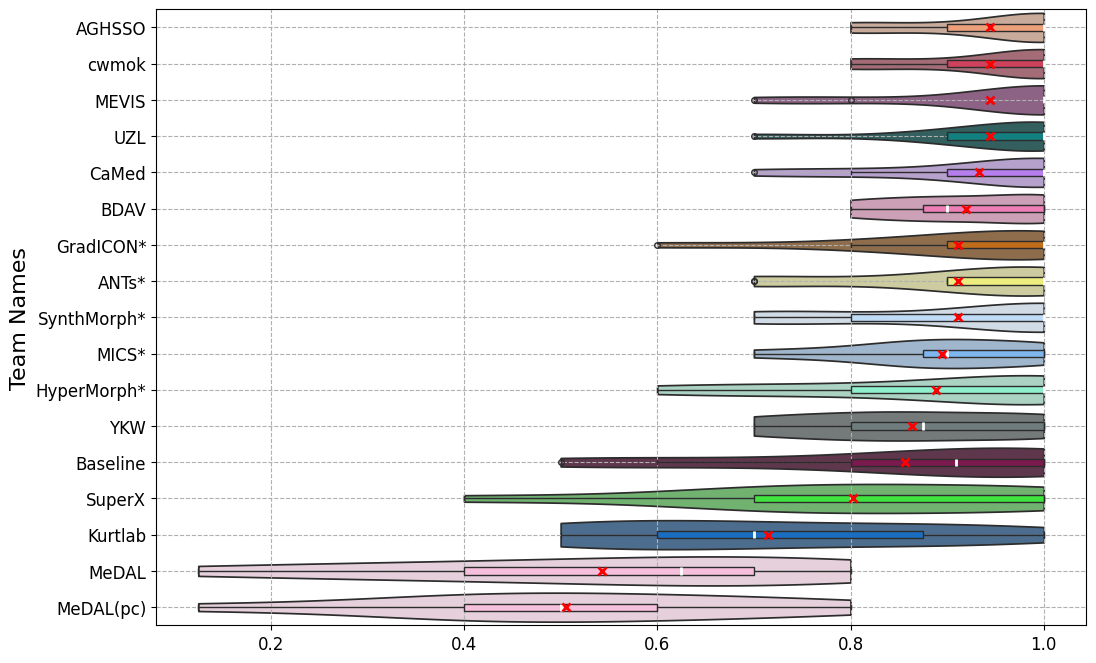}}
             \caption{Robustness for follow-up 2 to follow-up 1 to baseline registration task}
             \label{fig:f2 to f1 to b R}
         \end{subfigure}
        \caption{Comparative performance analysis of various participating methods in terms of Robustness on longitudinal data.}
         \label{fig:longitudinal analysis R}
      \end{figure}

      \begin{figure}[h]
         \centering
         \begin{subfigure}[b]{0.49\textwidth}
             {\includegraphics[width=\textwidth]{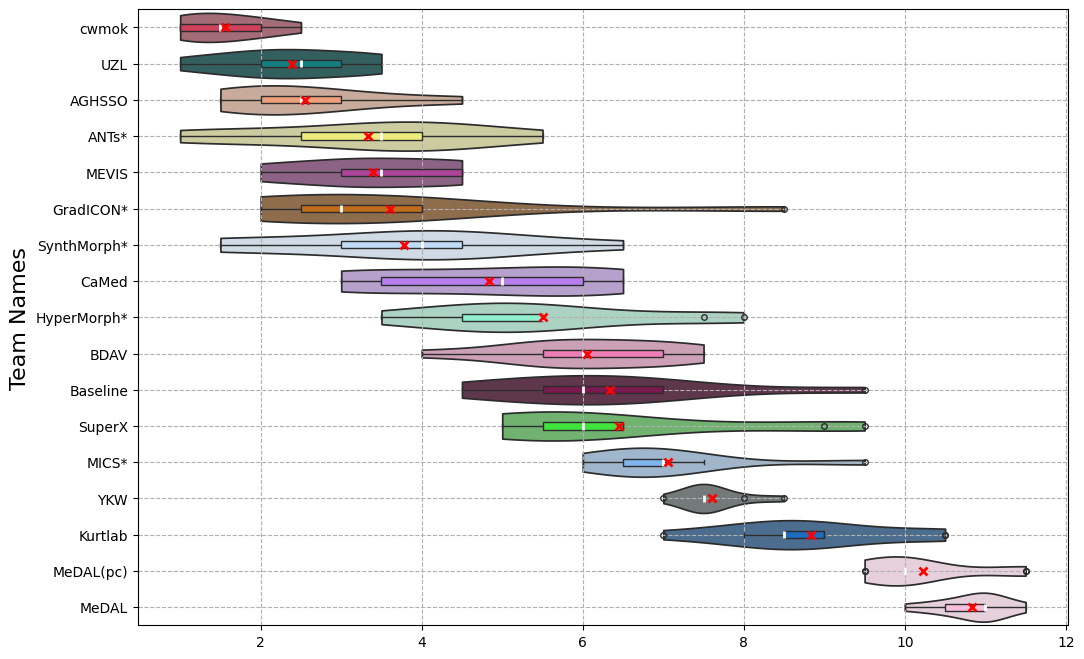}}
             \caption{BraTS-Reg score for follow-up 1 to baseline registration task}
             \label{fig:f1 to b BraTSReg score}
         \end{subfigure}
         \hspace{1mm}
         \begin{subfigure}[b]{0.49\textwidth}
             {\includegraphics[width=\textwidth]{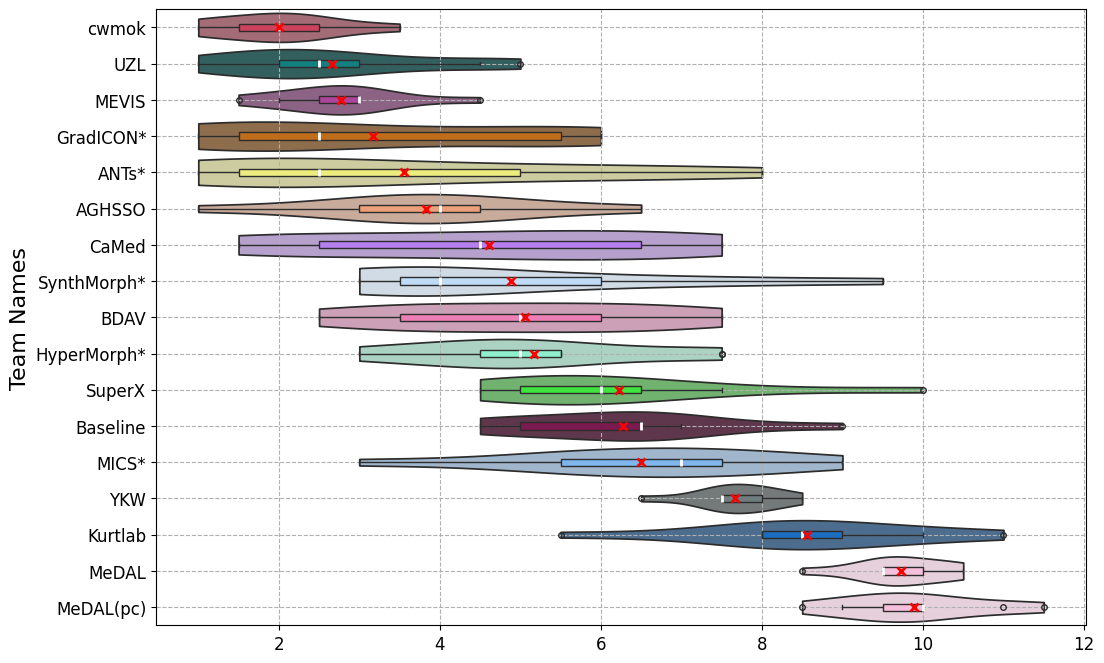}}
             \caption{BraTS-Reg score for follow-up 2 to baseline registration task}
             \label{fig:f2 to b BraTSReg score}
         \end{subfigure}
         \hspace{1mm}
         \begin{subfigure}[b]{0.49\textwidth}
             {\includegraphics[width=\textwidth]{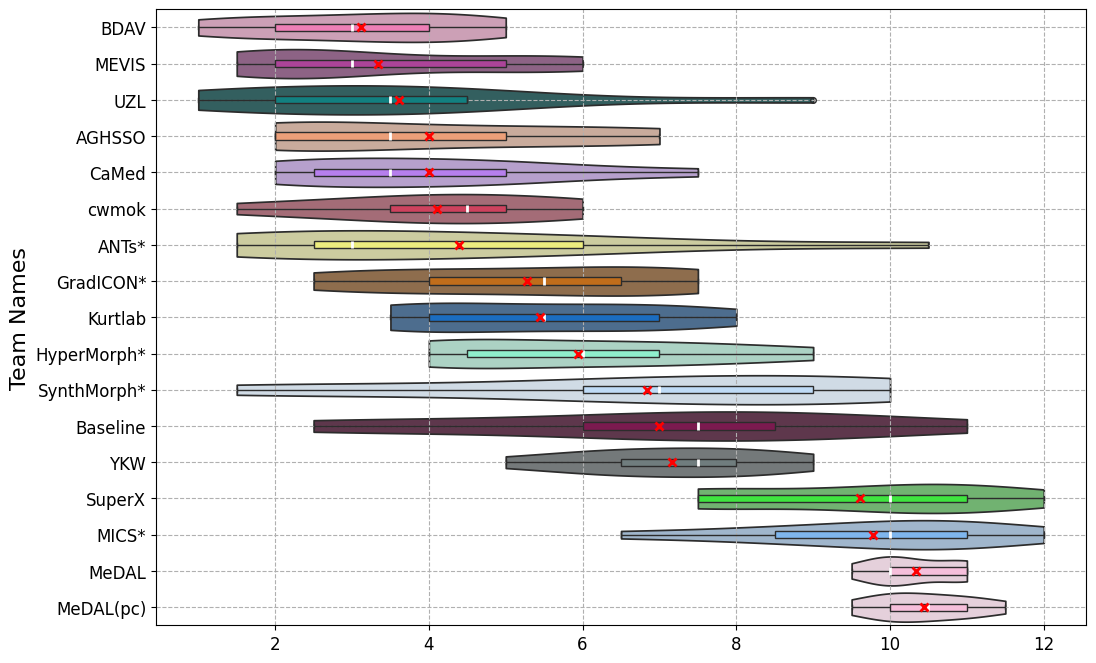}}
             \caption{BraTS-Reg score for follow-up 2 to follow-up 1 registration task}
             \label{fig:f2 to f1 BraTSReg score}
         \end{subfigure}
         \hspace{1mm}
         \begin{subfigure}[b]{0.49\textwidth}
             {\includegraphics[width=\textwidth]{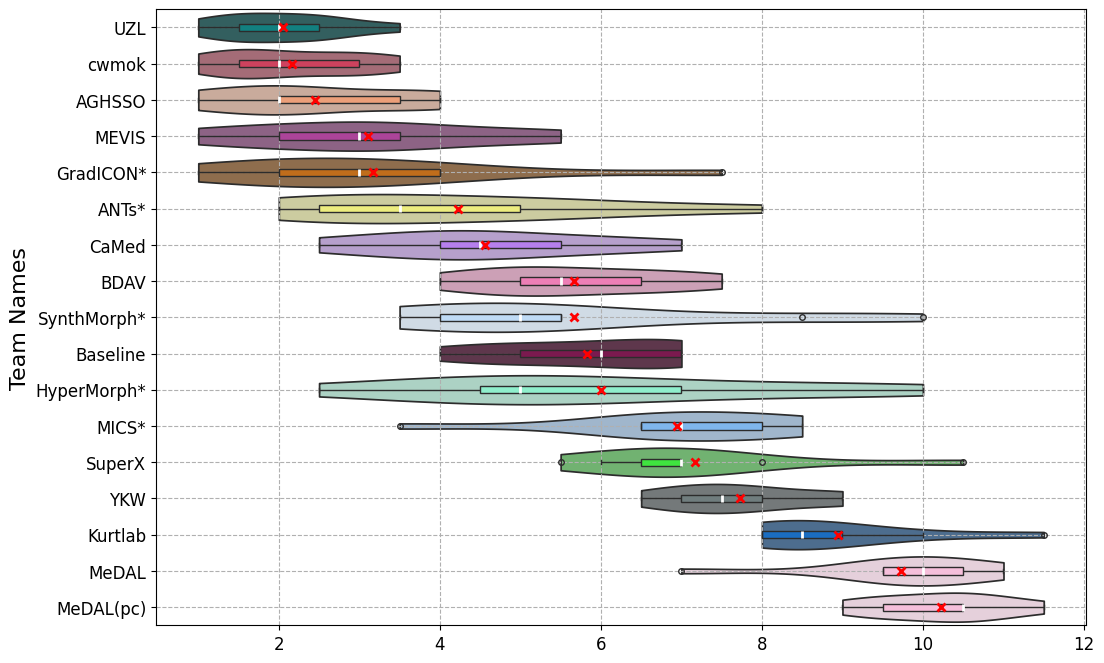}}
             \caption{BraTS-Reg score for follow-up 2 to follow-up 1 to baseline registration task}
             \label{fig:f2 to f1 to b BraTSReg score}
         \end{subfigure}
         \caption{Comparative performance analysis of various participating methods in terms of BraTS-Reg score on longitudinal data.}
         \label{fig:longitudinal analysis score}
      \end{figure}
The results of the longitudinal analysis in terms of MEE, Robustness, and the BraTS-Reg score are provided in Figures \ref{fig:longitudinal analysis MAE}, \ref{fig:longitudinal analysis R} and \ref{fig:longitudinal analysis score}, respectively. 
Comparing the overall ranges of the different metrics across the four tasks, regardless of methods used, we observed notable similarities in the MEE ranges for tasks 1, 2, and 4 (4.402±3.863, 4.547±3.498, 4.631±3.429, respectively). 
However, the MEE range was significantly smaller for task 3 (2.155±0.872). 
This may be because task 3 was the only task where registration was performed between two follow-up time points. 
The presence of similar deformations and structures in these scans likely rendered the registration between these two time points comparatively easier than the other three tasks. 
On the other hand, the overall robustness across all methods was generally lower in task 3 (0.543±0.277) compared to tasks 1, 2, and 4 (0.815±0.289, 0.848±0.196, 0.849±0.191 respectively). 
The very similar value ranges of all three metrics for task 2 and task 4 signify that there was no additional error introduced in deforming follow-up 2 to baseline scan when an intermediate registration step to follow-up 1 scan was involved.

In the analysis of the top-ranking teams for the four tasks, we noticed that in terms of MEE, team \emph{UZL} was the only method to be among the top 3 for all four tasks. 
Team \emph{cwmok} also performed well and was among the top 3 for all tasks except task 3. In terms of robustness, teams \emph{cwmok} and \emph{MEVIS} were consistently within the top 3 for tasks 1 to 4. 
In terms of BraTS-Reg score, the top ranking teams were similar to those in terms of MEE with team \emph{UZL} being the only method among the top 3 for all four tasks and team \emph{cwmok} being among the top 3 for all tasks, except task 3. 
Despite performing the best in terms of Robustness, team \emph{cwmok} dropped in ranking for task 3 in terms of MEE and hence BraTS-Reg score. 
This difference in results is possibly due to the inherent difference between task 3 and tasks 1, 2, 4 (i.e., follow-up to follow-up deformation instead of follow-up to baseline scans for which the algorithms were originally designed). 
In the analysis of the bottom-ranking teams, team \emph{MeDAL} consistently ranked among the bottom four in terms of all metrics for all tasks. 

\subsection{Analysis on the Relation of Performance and Inter-Rater Annotation Variability - Results}
\begin{figure}[!htbp]
         \centering
         \begin{subfigure}[b]{0.49\textwidth}
             {\includegraphics[width=\textwidth]{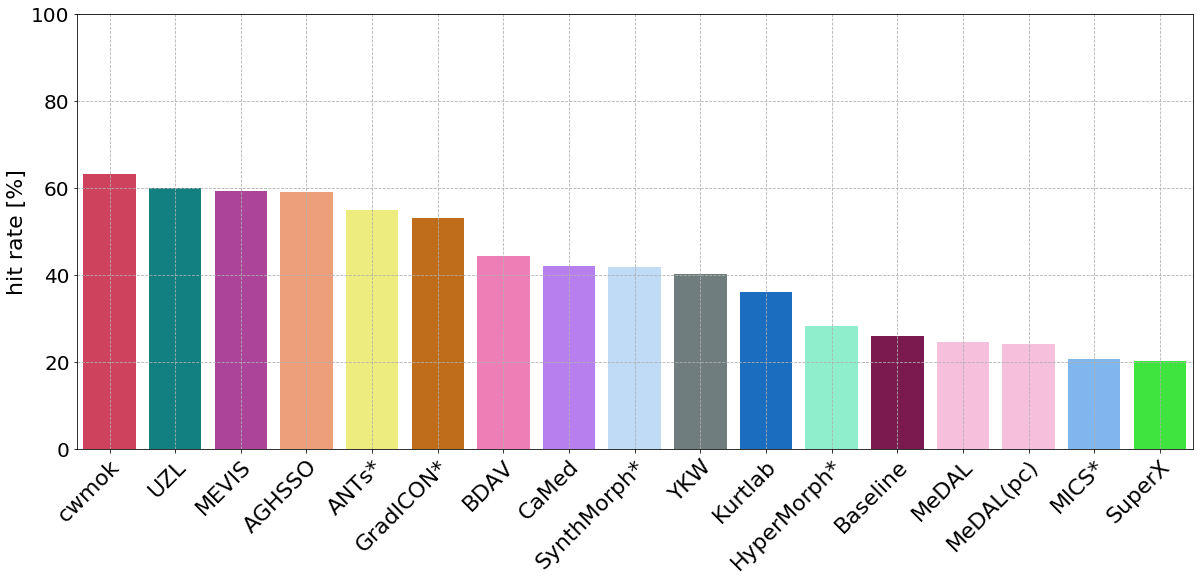}}
             \caption{\emph{Hit rate} per team on the respective landmarks' inter-rater annotation variability}
             \label{fig:hit_miss}
         \end{subfigure}
         \hspace{1mm}
         \begin{subfigure}[b]{0.49\textwidth}
             {\includegraphics[width=\textwidth]{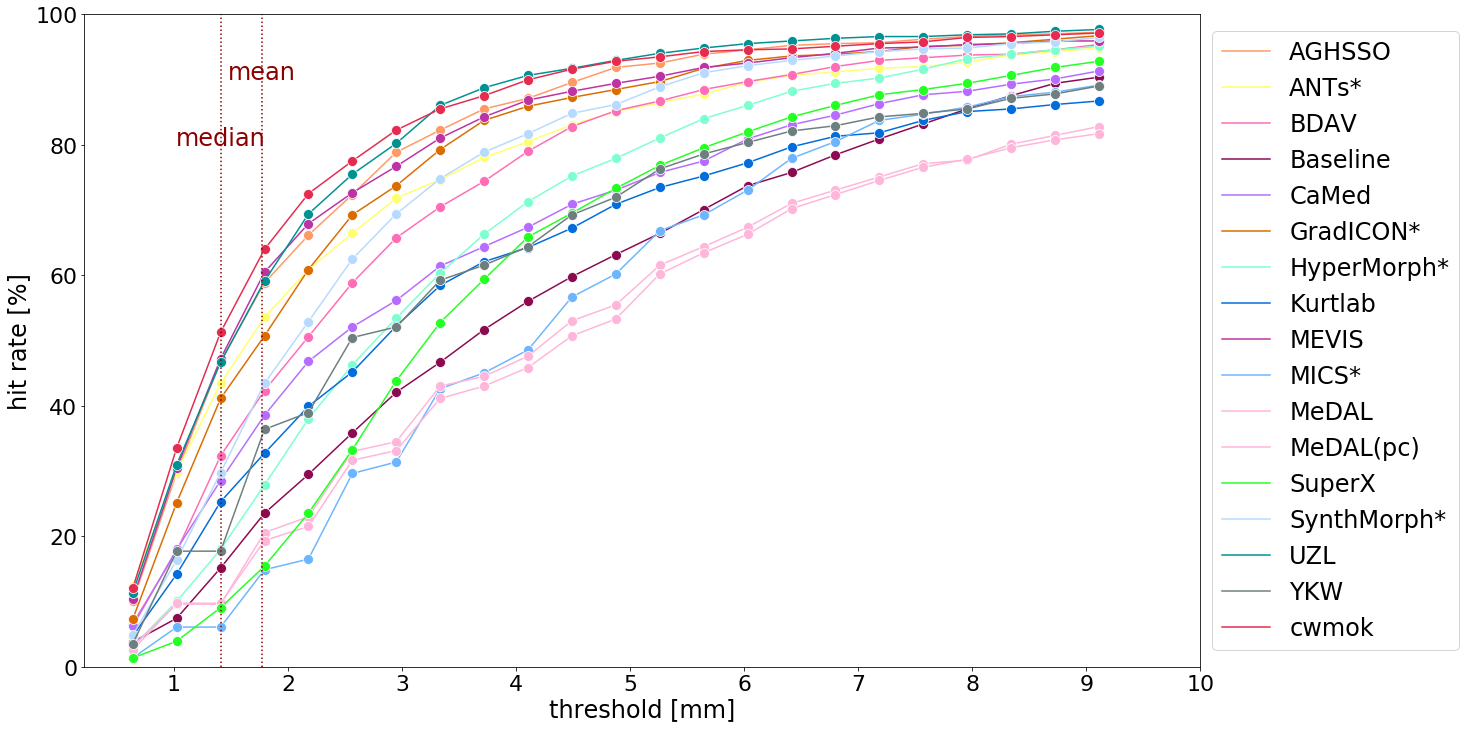}}
             \caption{\emph{Hit rate} per team on increasing thresholds}
             \label{fig:hits_lines}
         \end{subfigure}
         \caption{\emph{Hit rate} analysis.
         (a) gives an overview of the \emph{hit rate} per team using a threshold corresponding to the landmark-specific inter-rater annotation variability.
         (b) shows the \emph{hit rate} per team for increasing thresholds. Percentages in between evaluated thresholds are interpolated. Dotted lines indicate the mean and median inter-rater annotation variability.}
         \label{fig:hit_rate}
\end{figure}
We further evaluated the algorithms' results using \emph{hit rates}, as described in Section \ref{IR_methods_sec}. 
We compared the performance of all participating algorithms by calculating the \emph{hit rates} when being evaluated against the respective landmark-wise annotation variability. 
Since inter-rater analysis was performed on voxels with $1mm$ resolution, only landmarks with an AV of not less than $1mm$ ($81\%$) were considered for evaluation, in order to ensure a fair comparison against automated algorithms. 
From Fig. \ref{fig:hit_miss}, we can observe that the best-performing algorithm achieved \emph{hits} in around $60\%$ of all evaluated landmarks, showing that there is still room for improvement in terms of registration accuracy. 
Additionally, we evaluated the methods by increasing the error threshold in range $[0-10mm]$ and computing \emph{hit rate} curves, as illustrated in Fig. \ref{fig:hits_lines}. 
This provides an overview of the characteristics of the different algorithms under increasing error tolerance.

\section{Discussion and Conclusion}
\subsection{Identifying the Optimal Registration Algorithm}
It can be observed that there is some variation in ranking when comparing evaluation results on the test data vs. the OOD data. 
This might be explained by i) the origin of the respective data sets and ii) the differing experts annotating the two data sets, leading to possibly different difficulty levels. 
Nevertheless, we have a stable first rank with team \emph{cwmok}. Moreover, the methods ranked last just vary little among the different test sets and irrespective of the evaluation metric considered. 

\subsection{Establishing a Quantitative Baseline of State of the Art Algorithms}
Upon reviewing the methodologies of the top-performing teams, we observed that they included pre-alignment, deep neural networks, and inverse consistency analysis. 
Instance optimization on top of a deep learning approach, i.e., refining the registration result at test time in a post-processing step, also seems to be advantageous. 
Overall, the top-ranked methods were very close to each other in terms of all evaluation metrics. 
This is particularly evident in the case of the first three methods, and to a lesser extent, it remains true for the methods in the entire first half of the ranking. 

The first ranked method \emph{cwmok} achieved MEE equal to or even below respective inter-rater variability in about $60$\% of the evaluated landmarks. 
Therefore, there is still room for improvement in terms of accuracy and robustness, especially when being compared to human expert performance. While the majority of the evaluated algorithms were not developed for the very specific task of this challenge, there were few methods addressing the issue of missing correspondences directly by incorporating inverse consistency measurement (\emph{cwmok}, \emph{AGHSSO}). 
Only a handful of methods utilized the supplied landmarks for training, with none of them achieving top rankings. 
Despite the fact that the provided data had already been aligned to the same reference space, the application of rigid and/or affine pre-alignment appeared to enhance results, a practice employed by only a few methods prior to the actual deformable registration process. 

\subsection{Future Directions}
Throughout the process of conducting the first-ever BraTS-Reg challenge, considerable time and effort were invested in the annotation of landmarks. 
This task was distributed among various clinical experts, resulting in variations in the number and placement of landmarks for each individual case. While inter-rater analysis indicates a high level of agreement among experts in most cases, better generalizability may be achieved by offering a more comprehensive set of annotations. 
Also, the annotation protocol might benefit from stricter rules regarding the landmark locations, since the error measured at the annotations is biased by the fact that the landmarks are annotated at salient locations chosen by the clinical experts \citep{peter2021uncertainty}.
Furthermore, the evaluation process is likely to improve from the inclusion of supplementary qualitative metrics, such as assessing the smoothness of the displacement field or examining the algorithmic behavior in proximity to tumor-affected regions.


\section*{Acknowledgements}
 This study was supported in part by the the National Institutes of Health (NIH) under the awards from the 1) National Institute of Neurological Disorders and Stroke (NINDS) R01 grant (R01NS042645), 2) Informatics Technology for Cancer Research (ITCR) program of the National Cancer Institute (NCI) U24 grant (U24CA189523), 3) ITCR program of the NCI/NIH U01 grant (U01CA242871), 4) NIH Ruth L. Kirschstein Institutional National Research Service Award (T32EB001631), 5) RSNA Research \& Education Foundation under grant number RR2011, and 6) Deutsche Forschungsgemeinschaft (DFG) through TUM International Graduate School of Science and Engineering (IGSSE) GSC 81. The content of this publication is solely the responsibility of the authors and does not necessarily represent the official views of NIH, the RSNA R \& E Foundation, or any of the other funding bodies.
 B.Baheti, H.Akbari, and Spyridon Bakas conducted part of the work reported in this manuscript at their current affiliation, as well as while they were affiliated with the Center for Artificial Intelligence and Data Science for Integrated Diagnostics (AI2D) and the Center for Biomedical Image Computing and Analytics (CBICA) at the University of Pennsylvania, Philadelphia, PA 19104, USA. 

\bibliographystyle{model2-names.bst}\biboptions{authoryear}
\bibliography{main_arxiv}

\clearpage

\clearpage
\section*{Appendix A.}

\tikzbullet{amaranth}{amaranth} \hspace{2mm} \textbf{Team cwmok}\\
Mok and Chung \citep{mok2022robust} proposed a 3-step registration method, which comprises an affine pre-alignment, a convolutional neural network with forward-backward consistency constraint, and a nonlinear instance optimization. First, possible linear misalignments caused by the tumour mass effect were eliminated with the descent-based affine registration method. Second, conditional deep Laplacian pyramid image registration network with forward-backward consistency (DIRAC) \citep{mok2022unsupervised, mok2021conditional} was leveraged to jointly estimate the regions with missing correspondence and bidirectional nonlinear displacement fields for the pre-operative and follow-up scans. During the training phase, regions with missing correspondence were excluded from the similarity measure. This reduced the effect of the pathological regions on the registration algorithm in an unsupervised manner. Finally, non-rigid instance optimization with forward-backward consistency constraint was introduced to correct solutions from the previous step that  were biased because of insufficient training and discrepancy in distributions. This step further improved
the robustness and registration accuracy of initial solutions. The non-parametric deformation was controlled by the forward-backward consistency constraint as in the previous step and was updated using an Adam optimizer together with multi-level continuation to avoid local minima. The implementation of DIRAC is available at \url{https://github.com/cwmok/DIRAC}. 

\tikzbullet{Darkcyan}{Darkcyan} \hspace{2mm} \textbf{Team UZL}\\
UZL utilised a combination of hand-crafted features, a single-level discrete correlation layer subject to a convex optimisation scheme, and a subsequent Adam instance optimisation for fine-grained displacement prediction. First, MIND-SSC features were extracted on T1 contrast-enhanced baseline and follow-up images. The features were then used to compute a correlation-based cost tensor containing sum-of-squared-differences. Next, an iterative convex optimisation regarding feature matching and global smoothness was performed. To account for tumor-related mass effects and missing correspondences the authors employed a large search range and enforced inverse consistency. An Adam instance optimisation further refined the intermediate displacement field with diffusion regularisation and B-spline-interpolation. 

\tikzbullet{atomictangerine}{atomictangerine} \hspace{2mm} \textbf{Team AGHSSO}\\
The proposed method consists of: (i) preprocessing and modality aggregation, (ii) iterative affine registration, (iii) dense displacement field calculation by LapIRN, (iv) iterative, instance optimization-based nonrigid registration, and (v) displacement field fine-tuning by optimizing an objective function that was weighted based on the inverse consistency. The method addressed both the challenges related to the pre- to post- operative registration, namely the large, nonrigid deformations and the missing tissues. The large and complex deformations were addressed by the LapIRN network with large enough receptive field, while the missing tissues were handled by the proposed inverse consistency-based objective function weighting. The objective function (both the similarity measure and the regularization term) were weighted by the corresponding inverse consistency error during the fine-tuning step. As a result, the registration accuracy close to the missing tumor and its cavity was improved.

\tikzbullet{lightpink}{lightpink} \hspace{2mm} \textbf{Team BDAV USYD}\\
This team adopted the recently proposed Non-Iterative Coarse-to-fine registration Network (NICE-Net) \citep{meng2022non} as the backbone and extended it by introducing dual deep supervision. The NICE-Net consists of a feature learning encoder and a coarse-to-fine registration decoder. The feature learning encoder has two identical, weight-shared paths to extract features from the fixed and moving images separately, which  were then propagated to the coarse-to-fine registration decoder. The decoder performs multiple steps of coarse-to-fine registration in a single network iteration. Dual deep supervision, including a deep self-supervised loss based on image similarity (local normalized cross-correlation) and a deep weakly-supervised loss based on manually annotated landmarks (mean square error), was embedded into the NICE-Net (referred as NICE-Net-ds). As the provided training set was relatively small (140 intra-patient image pairs), the NICE-Net-ds was first pretrained with inter-patient image pairs (280 $\times$ 279 pairs) to avoid overfitting. Then, the NICE-Net-ds was further trained for intra-patient registration with dual deep supervision. During inference, pair-specific fine-tuning was performed to improve the network’s adaptability to testing variations. In addition, as the MRI scans provided by the challenge organizers had been rigidly registered to the same anatomical template, this method solely optimized for deformable image registration without considering affine registration. 

\tikzbullet{lightviolet}{lightviolet} \hspace{2mm} \textbf{Team CaMed}\\
 An enhanced unsupervised learning-based method was developed for reliable and accurate registration of patient-specific brain MRI scans containing pathologies.  The proposed method extends our previously developed registration framework iRegNet \citep{9570282}. In particular, incorporating an unsupervised learning-based paradigm as well as several minor modifications to the network pipeline, allowed the enhanced iRegNet method to achieve respectable results. Similar to the baseline iRegNet, the registration procedure consists of two steps: First, $I_{B}$ and $I_{F}$  were fed into our convolutional neural network (CNN) that then predicts $\phi$. Second, $I_{F}$ was transformed into a warped image ($I_{F}. \phi$) using a spatial re-sampler. 

\tikzbullet{pureblue}{pureblue} \hspace{2mm} \textbf{Team Kurtlab}\\
 A two-stage cascaded network was developed consisting of the Inception and the TransMorph architecture. A series of Inception modules were initially used to fuse the 4 image modalities inputs and extract their most relevant information. In short, the Inception module was used to process each contrast separately, extracting the relevant information before concatenating them. The concatenated data was then passed through more Inception modules that merged the contrasts together and output new moving and target images. This approach had several advantages. First, it added more training parameters corresponding to the data merging layers. It also helped reduce the memory requirements by merging 4 volumetric images into a single one. The output of the Inception model was then sent into a variant of the TransMorph architecture to generate displacement fields for transforming the post-surgery images to their corresponding pre-surgery ones. TransMorph, a hybrid TransformerConvNet, was able to determine which parts of the input sequence  were essential based on contextual information through the use of self-attention mechanisms. Finally, the loss function was composed of a standard image similarity measure and a diffusion regularizer.

\tikzbullet{aurometalsaurus}{aurometalsaurus} \hspace{2mm} \textbf{Team YKW}\\
The QPDIR algorithm was an intensity-based algorithm, which transforms the computation of deformation field to an optimization problem aimed at minimizing terms related to image dissimilarity and regularization. The terms were computed based on performing an exhaustive search among image blocks. The optimization was performed using a gradient-free quadratic penalty method. The whole optimization problem was decomposed to several sub-problems and each of them can be solved by straightforward block coordinate decent iteration. The QPDIR algorithm consisted of 3 steps. Firstly, the objective function was formulated by combining an image dissimilarity term and a regularization term determined through exhaustive search. The gradient-free quadratic penalty method was then employed to optimize the objective function. Next, the search window size was gradually reduced, and step 1 was repeated until the search window size reached a sufficiently small scale. Finally, the full displacement field was computed by applying moving least square (MLS). It's worth noting that multi-modal registration was achieved by fusing the results of single-modality registration.

\tikzbullet{palepink}{palepink} \hspace{2mm} \textbf{Team MeDAL} \hspace{2mm} and \hspace{2mm} MeDAL(pc)\\
The proposed method consists of two stages, a segmentation stage, and a subsequent registration stage.  The segmentation stage consists of two U-Nets with shared parameters \citep{schwarz2007non}. The U-Nets  were similar, with each one of them containing three levels with residual blocks at every level and the number of feature maps starts from 8. Each U-Net segments the regions of interests (ROI), which are patches of sizes ($9\times9\times9$) around the landmarks of the input volume. These landmarks can be found in either the moving volume or the fixed volume. The segmentation network was followed by an attention block in which the output of the U-Net (a binary segmentation map) was multiplied by the input volume to produce an attentive volume. The concatenated outputs (the attentive fixed and moving volumes) of the segmentation network serve as an input to the registration network. To tackle the problem of class imbalance between the foreground and the background of the segmentation mask, we used the focal loss between the segmentation masks and the predicted segmentation maps \citep{machado2018deformable}. The architecture of the registration network was the same as the U-Net architecture used for segmentation. The difference was that the network for registration outputs a deformation field instead of a segmentation map. The deformation field was used for deforming the moving volume to match the fixed one. The loss function of the registration network was a combination of two losses, the similarity loss, and the smoothness loss. \\
\textbf{MeDAL(pc)}: In the original challenge submission, the team had problems with loading the landmarks properly. After fixing the issues, the team was allowed to provide a post-challenge submission of their method, indicated with \emph{(pc)}.

\tikzbullet{byzantine}{byzantine} \hspace{2mm} \textbf{Team Fraunhofer MEVIS}\\
The proposed method was an iterative variational image registration approach based on \citep{modersitzki2009fair}, in which the registration of two volumes can be modeled as the minimization of a discretized objective function. The final solution consists of a parametric and a non-parametric step. In the parametric step, the registration task was based on the estimation of a limited number of transformation parameters. In particular, a rigid registration, restricted to the search of rotations and translations parameters, was computed. In the parametric approach, the objective function only consists of the distance measure, computed between the fixed (post-operative) and warped moving (pre-operative) image. As a distance measure, we utilized the normalized gradient field (NGF) \citep{haber2006intensity}. The transformation matrix obtained in the parametric registration initializes the non-parametric step, in which a deformation vector field was computed. In the deformable solution, the NGF was also utilized as a distance measure. To limit the possible registration solutions and make the deformation field more plausible, two regularization terms were added to the objective function. The first one was the curvature regularizer \citep{fischer2003curvature}, which penalizes deformation fields having large second derivatives. Additionally, the volume change control was utilized to reduce foldings in the deformation field \citep{ruhaak2017estimation}. In the parametric and non-parametric steps, the registration was conducted on three levels using images at different scales. The deformation was initially computed on the coarsest level, where the images  were downsampled by a factor equal to $2^{(L-1)}$. On a finer level, the previously computed deformations  were utilized as an initial guess by warping the moving image. At each level, the moving and fixed images  were downsampled. Furthermore, the choice of the optimal transformation parameters was conducted by using the quasi-Newton l-BGFS, due to its speed and memory efficiency \citep{liu1989limited}.
Each step of the proposed registration method can process only one MRI sequence at a time. Thus, we first verified on the training set of the BraTS-Reg challenge dataset which sequence was the best to guide the registration in both the rigid and non-rigid approaches. Our final solution used the T2 images in the parametric step, whereas the T1c acquisitions of the corresponding volumes guided the deformable registration. 

\tikzbullet{limegreen}{limegreen} \hspace{2mm} \textbf{Team SuperX}\\
This method consists of two steps i) A rigid registration method, the Nelder-Mead method (also named downhill simplex method) and ii) Affine transformation for rigid registration algorithm which was applied on floating image before non-rigid registration was performed using free-formed deformations. In the non-rigid registration step, two optimization methods were investigated. The first one was particle swarm optimization, which is a computational method that optimizes a problem by iteratively improving a candidate solution according to  a given measure of quality. The second optimization method that was examined was the downhill simplex method which is a commonly used local fast optimization. According to our experimental results, these methods achieved similar accuracy. We finally used the downhill simplex method to minimize the correlation coefficient as an image similarity function by downhill simplex optimization.  In the non-rigid image registration, we used free-form deformation, where the force exerted on each floating voxel drives it to the correct position to match the reference volume. Our algorithm used scaling factors of 0.125, 0.25, and 0.5 to generate a multi-level pyramid and accelerate the running time. We initially optimized at a lower resolution level and then scaled up the warp field from lower to higher resolutions. According to Information Theoretic Similarity Measures in Non-Rigid Registration, the force field was based on joint entropies. 

\tikzbullet{lightyellow}{lightyellow} \hspace{2mm} \textbf{Team ANTs\textsuperscript{*}}\\
The BraTS-Reg22 data were processed using previously vetted and frequently used registration parameter sets \citep{avants2011reproducible,avants2014insight}, which have been packaged within the different ANTsX platforms, specifically ANTs, ANTsPy, and ANTsR (\url{https://github.com/ANTsX}). Each of these parameter sets consists of multiple transformation stages for determining anatomical correspondence. Initially, linear transform parameters are estimated, including center of (intensity) mass alignment followed by optimization of both rigid and affine transforms using mutual information as the similarity metric \citep{viola1997alignment}. The final deformable alignment utilized symmetric normalization (SyN) with Gaussian \citep{avants2008symmetric} or B-spline \citep{tustison2013explicit} regularization and neighborhood cross-correlation \citep{avants2008symmetric} or mutual information similarity metric. The effects of image modality choice including all single modalities and combinations of modality pairs (specifically, T1-contrast enhanced/T2 and T1-contrast enhanced/FLAIR) were also explored. Although performance with the training data was similar across the different configurations, SyN with Gaussian regularization and neighborhood cross-correlation (radius = 2 voxels) using T1-contrast enhanced images was selected for a single submission during the validation phase. Further details on this internal evaluation, including the precise ANTsPy calls, can be found at a dedicated GitHub repository (\url{https://github.com/ntustison/BraTS-Reg22}).

\tikzbullet{paleblue}{paleblue} \hspace{2mm} \textbf{SynthMorph\textsuperscript{*}} \hspace{2mm} and \hspace{2mm} \tikzbullet{aquamarine}{aquamarine} \hspace{2mm} \textbf{HyperMorph\textsuperscript{*}}\\
\noindent These approaches  were build on VoxelMorph~\citep{balakrishnan2019}, a widely-used learning-based framework for pairwise deformable registration that aligns a moving image \emph{m} with a fixed image \emph{f} by predicting a dense correspondence~$\phi$. This framework leverages a convolutional architecture~$g_\theta$, with trainable parameters~$\theta$, that takes as input~$\{m,f\}$ and outputs a stationary velocity field, integrated via squaring-and-scaling~\citep{dalca2018} to yield the diffeomorphic map~$\phi_\theta$. VoxelMorph generally optimizes a loss combining an image matching term~$\mathcal{L}_{sim}$ with a regularization term~$\mathcal{L}_{reg}$ to encourage smooth deformations:
\begin{equation}
\nonumber
\mathcal{L}(m, f, \phi_\theta)
= \mathcal{L}_{sim}(m \circ \phi_{\theta}, f)
+ \lambda \mathcal{L}_{reg}(\phi_\theta),
\end{equation}
where~$m \circ \phi_{\theta}$ represents~$m$ transformed by~$\phi_\theta$, and~$\lambda$ is a regularization weight. In this challenge, we used normalized cross-correlation for~$\mathcal{L}_{sim}$ and defined ~$\mathcal{L}_{reg}(\phi)=\frac{1}{2}||{\nabla}u||^2$, where $u$ was the displacement of deformation~$\phi$. Network $g_\theta$ implements a U-Net~\citep{ronneberger2015}, using convolutional kernels of size~$3^3$ and Leaky-ReLU activations. We treat~$m$ as the baseline and~$f$ as the follow-up scan for each subject, stacking all four contrasts to compose multi-channel input images.

VoxelMorph generally performs better on affinely aligned images and, like most learning-based approaches, does not generalize well to modalities outside the training domain. Furthermore, registration quality was often sensitive to the selected values of hyperparameters, such as~$\lambda$. To address this set of challenges, two methods that implement and build on the VoxelMorph framework  were evaluated: SynthMorph and HyperMorph.

\vspace{6pt}

\noindent \textbf{SynthMorph} is a fully-convolutional architecture for joint affine and deformable registration trained on a wide range of synthetic brain images, enabling the network to generalize across MRI contrasts, resolutions, and anatomies~\citep{hoffmann2022synthmorph}. First, an affine network $h_\eta$, with parameters $\eta$, predicted separate feature maps for~$m$ and $f$, and aligned the barycenters of these maps using a least-squares regression, yielding a linear transform~$T_\eta = h_\eta(m, f)$ that was applied to~$m$~\citep{hoffmann2023affine}. Then, network~$g_\theta$ was used to predict the deformation field~$\phi_\theta = g_\theta(T_\eta \circ m, f)$.

A pre-trained SynthMorph model was fine-tuned to the challenge data using a semi-supervised approach, optimizing an additional loss term~$\mathcal{L}_{sup}$ that measures the mean squared error between moving and fixed landmarks $\{x_m, x_f\}$:
\begin{equation}
\nonumber
    \mathcal{L}(T_\eta, \phi_\theta, m, f, x_m, x_f)
    = \mathcal{L}_{sim}(m \circ T_\eta \circ \phi_\theta, f)
    + \lambda_1 \mathcal{L}_{reg}(\phi_\theta)
    + \lambda_2 \mathcal{L}_{sup}(x_m, x_f),
\end{equation}
with~$\lambda_1 = 32$ and~$\lambda_2 = 1$. The networks~$h_\eta$ and~$g_\theta$  were initialized with pre-trained weights and the affine component~$h_\eta$ before fine-tuning $g_\theta$. The model comprised 20 total convolutions of 256 channels each. During training a learning rate of~$10^{-5}$ was used and data augmentation was applied following~\citet{hoffmann2023joint}.

\vspace{6pt}
\noindent \textbf{HyperMorph} facilitates training a \emph{single} model encompassing a landscape of possible values for~$\lambda$~\citep{hoopes2021hypermorph,hoopes2022}. This avoids the need to train several separate models and allows for fine-scale hyperparameter choice during inference. In HyperMorph, a hypernetwork learns to model the effect of varying~$\lambda$ to predict the corresponding parameters of~$g_\theta$.

Using a subset of the challenge training data, first HyperMorph was trained and the effect of~$\lambda$ on a held-out validation set was evaluated. Based on a combination of evaluating both visual deformation quality and distance between annotated landmarks, $\lambda = 2.4$ was selected and the base VoxelMorph model was trained until convergence. The pre-trained affine component of SynthMorph was employed as a pre-processing step. The model~$g_\theta$ comprised 12 total convolutions, each with 64 channels, and was trained with a learning rate of~$10^{-4}$, using the augmentation strategy noted above.

\tikzbullet{pureorange}{pureorange} \hspace{2mm} \textbf{Team GradICON\textsuperscript{*}}\\
GradICON’s training protocol and hyperparameters \citep{tian2022gradicon}  were adopted. Its ability to generalize was assessed by investigating its performance without explicitly modeling image differences due to tumor resection with two significant changes in the original approach. Initially, the number of input channels of the first convolutional layer was increased to match the number of modalities. This adjustment enabled the utilization of visual cues across different modalities. The image similarity was computed by defining the  local normalized cross correlation (LNCC) as an average over the LNCCs for each modality (channel). The second modification consisted of a new training strategy to alleviate overfitting caused by the small available training dataset. The network was pre-trained following the original training process in \citep{tian2022gradicon} with inter-patient pairs from the train set. Subsequently, the entire network (Stage1 and Stage2) was fine-tuned using intra-patient pairs from the train set. The input images  were normalized to [0, 1] per modality. In the pre-train phase, random pairs of pre-operative and follow-up images  were picked as training pairs. In the fine-tune phase, the paired images provided by the challenge  were used. 

In both phases, the network was trained with the GradICON’s default hyperparameters (i.e., learning rate of $5e^{-5}$, regularizer weight $\lambda = 1.5$, and ADAM optimizer). Due to memory limitations, a batch size of 2 was used when training Stage 1 \citep{tian2022gradicon}. A batch size of 1 was used for Stage 2 \citep{tian2022gradicon} in the inter-patient (pre-training) phase and in the intra-patient training (fine-tuning) phase. The network was trained for 20,000
iterations for pre-training and 10,000 iterations for fine-tuning. The protocol described in \citep{tian2022gradicon} was followed for inference where a displacement field was directly predicted using the trained network from the input images without any pre-alignment. Subsequently, 50 iterations of instance optimization  were used. Due to memory limitations, this optimisation was only over Stage 2 with a learning rate of $5e^{-5}$ keeping Stage 1 frozen. The implementation was available at \url{https://github.com/uncbiag/ICON}.

\tikzbullet{lightblue}{lightblue} \hspace{2mm} \textbf{Team MICS\textsuperscript{*}}\\
The method was based on the multi-task registration framework presented in \citep{estienne2020deep}. In summary, the architecture utilized a shared UNet-like encoder that maps both inputs (moving $M$ and reference $R$) to a latent code using a late fusion strategy. The final layer of a UNet-like registration decoder, was based on the formulation presented in \citep{stergios2018linear}. Specifically, a cumulative sum across all dimensions was performed on the output of a logistic growth function (i.e., $f(x) = L / 1 + e^{-k(x-x_0)}$, with  $L=8$, $k=1$, $x_0=1$) that forces the values to be strictly positive an on a specific range. By enforcing these displacements to have positive values and subsequently applying an integration operation along each dimension, the spatial sampling coordinates can be retrieved. Such an approach ensures the generation of smooth deformations that avoid self-crossings, while allowed the control of maximum displacements among consecutive pixels using appropriate parameterization of the logistic growth function. The training was performed in two steps. In the first step a random pairing scheme was utilized to generate different input pairs drawn from different cases. In the second step, the model that was derived was fine-tuned only on baseline-follow-up pairs for each patient individually. The losses that were utilized for these two training steps  were $L1$ and $L2$ respectively:

\begin{equation}
\nonumber
\begin{split}
L1 & = L_{rec} + L_{ncc} + L_{jac} + L_{cycl}\\
L2 & = L_{rec} + L_{ncc} + L_{jac} + L_{cycl} + L_{lndmrk}
\end{split}
\end{equation}
where $L_{rec}$ was the mean squared error between the target and deformed image, $L_{ncc}$ was the normalized cross correlation loss between the target and deformed image, $L_{jac}$ was the Jacobian determinant of the deformation field, $L_{lndmrk}$ was the mean absolute error between the deformed and target landmark locations and lastly, $L_{cycl}$ was the mean squared error between the moving image and the output of a sequential application of a forward and backward deformation. It should be noted that, in both training stages a cycling scheme was used where all the aforementioned losses  were averaged by considering both inputs as moving/reference images, respectively.


\end{document}